\documentclass[review]{elsarticle}
\usepackage{amsmath}
\usepackage{url}
\usepackage{lineno}
\usepackage{hyperref}
\usepackage{amssymb}
\usepackage{graphicx}
\usepackage{epsfig}
\usepackage{epstopdf}
\usepackage{booktabs}
\usepackage{multirow}
\usepackage{color}
\usepackage{multicol}
\usepackage{amsfonts}
\usepackage{textcomp}
\usepackage{changepage}
\usepackage{algorithm} 
\usepackage{algorithmic}
\usepackage{geometry}
\usepackage{subfigure}

\modulolinenumbers[5]
\journal{Journal of \LaTeX\ Templates}
\usepackage{array}

\biboptions{sort&compress}

\begin{document}

\begin{frontmatter}

\title{Recent Advances and New Guidelines on Hyperspectral and Multispectral Image Fusion
}

\author{Renwei Dian, Shutao~Li$^*$,  Bin Sun, and Anjing Guo }
\address{College of Electrical and Information Engineering, Hunan University, Changsha, China}





\begin{abstract}
Hyperspectral image (HSI) with high spectral resolution often suffers from low spatial resolution owing to the limitations of imaging sensors. Image fusion is an effective and economical way to enhance the spatial resolution of  HSI, which combines  HSI with higher spatial resolution multispectral image (MSI) of the same scenario. In the past years, many HSI and MSI fusion algorithms are introduced to obtain high-resolution HSI. However, it lacks a full-scale review for the newly proposed HSI and MSI fusion approaches. To tackle this problem,
  this work gives a comprehensive review and new guidelines for  HSI-MSI  fusion. According to the characteristics of HSI-MSI fusion methods, they are categorized as four categories,  including pan-sharpening based approaches, matrix factorization based approaches, tensor representation based approaches, and deep convolution neural network based approaches.
  We make a detailed introduction, discussions, and comparison for the fusion methods in each category. Additionally,  the existing challenges and possible future directions for the HSI-MSI  fusion are presented.
\end{abstract}

\begin{keyword}
  Hyperspectral and multispectral image fusion  \sep Hyperspectral image super-resolution \sep Hyperspectral imaging
\end{keyword}

\end{frontmatter}

\section{Introduction}
\begin{figure*}[t]
\begin{center}
{\includegraphics[width=150mm]{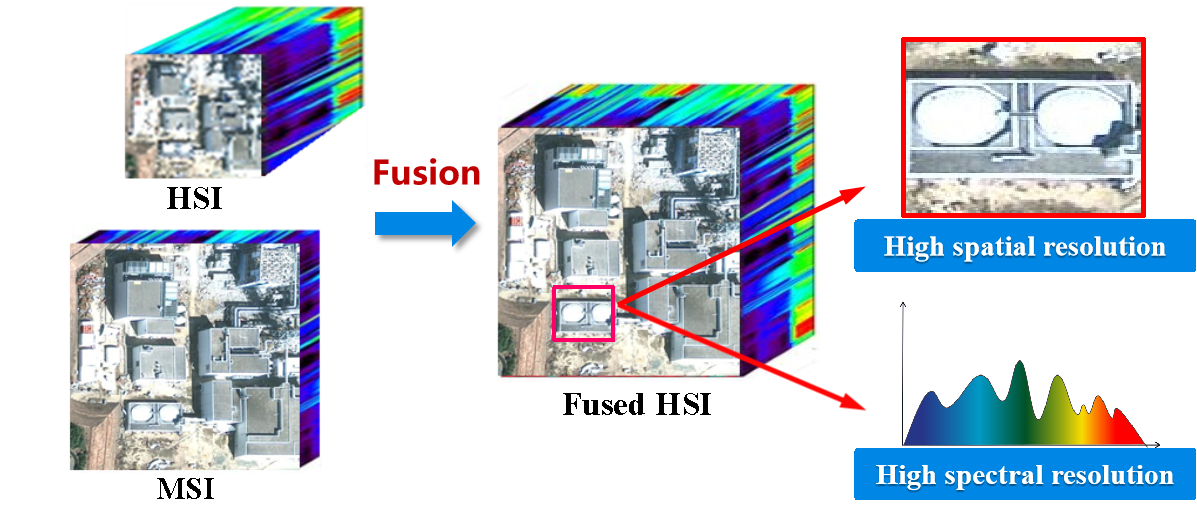}}
\end{center}
   \caption{  The illustration of the HSI-MSI fusion. }
\label{fig1}
\end{figure*}

Hyperspectral imaging sensor can collect dozens of
or hundreds of spectral bands in wide range spectral
coverage. Since  the materials often have different reflectance
 for different wavelength, hyperspectral image (HSI) enables
  accurate identification of materials owing to its high
 spectral resolution and wide spectral range. In this way,
  hyperspectral imaging has found comprehensive applications
   on remote sensing \cite{1}, face recognition \cite{Pan2003Face},
      medical diagnosis \cite{medicalimage},  etc.
     However,  on account of  limitations for imaging
     cameras, there is the certain tradeoff for the spectral
     resolution and spatial resolution. Hence, HSI with
      a large number of bands  usually has a low spatial resolution
      to ensure high SNR. On the contrary,  imaging sensors
      can obtain an image with a higher spatial resolution but
       with a small number of spectral bands, consisting of
        RGB image, panchromatic image,  and multispectral image (MSI).
  As shown in Table \ref{sensors}, spaceborne imaging sensor can acquire the high-resolution panchromatic image and four-band MSI
    with decimetric spatial resolution and
metric spatial resolution, respectively. However,
 HSI can only be acquired with a spatial resolution of dozens of meters, which hinders the applications of HSI. One effectual and economical way to improve the spatial resolution of HSI is image fusion.
  More and more sensors can simultaneously acquire HSI
  and high-resolution MSI  on the same scenario,
   and therefore image fusion approaches  can be performed
   to acquire a fused  image with the high spectral resolution
    and  high spatial resolution, which is referred to as HSI-MSI fusion.
  For example,   Hyperspectral
 imager suite (HISUI), the Japanese
next-generation earth-observing sensor, simultaneity has
 hyperspectral and multispectral imaging sensor,
  where the  spatial resolution of HSI and MSI  are 30m and 5m,
  respectively. Gaofen (GF)-5 satellite and Gaofen (GF)-2 satellite,
   designed by China, takes the hyperspectral imaging and multispectral
   imaging sensors, respectively, where the HSI and MSI have the ground
    sampling distance (GSD) of 30m and 4m, respectively.
     As illustrated in Figure \ref{fig1}, HSI and MSI fusion is
      an effective  and common way to improve the spatial
     resolution of the HSI.  The HSI and MSI fusion is the part of
      pixel-level image
     fusion \cite{681,691,701,711,721,LIU2018158,LI2017100,liu2020remote,meng2019review,751}.
The fused image of high spatial  and spectral resolution can
 help us better recognize and understand the materials,
  which has contributed  to many tasks including
  object classification \cite{82}, anomaly detection \cite{81},  change detection \cite{80}, etc.  Reference \cite{81} shows that  the fused HSI  indeed improves the detection accuracy compared with the original HSI.
   The fusion technique can find application in  high spatial resolution ecosystem monitoring,  minerals survey, plant investigation,  and disaster warning.


\begin{table*}[t]
\begin{center}
\caption{  The ground sampling distance and spectral resolution for some spaceborne imaging sensors }\label{sensors}
\vspace{2mm}
\begin{tabular}{ccccc}
\toprule
Sensor & \multicolumn{2}{c}{Spectral range (nm)}                                      & Number of bands & GSD (m) \\
 \midrule
\multirow{2}{*}{HYPXIM} & \begin{tabular}[c]{@{}c@{}}HSI\end{tabular}          & 400-2500       & 210  & 8     \\
                        & \begin{tabular}[c]{@{}c@{}}PAN \end{tabular}           & 400-800       & 1    & 2     \\
\multirow{2}{*}{HISUI}  & \begin{tabular}[c]{@{}c@{}}HSI \end{tabular}          & 400-2500   & 185  & 30    \\
                        & \begin{tabular}[c]{@{}c@{}}MSI\end{tabular}          & 450-900    & 4    & 5     \\
{GF-5}   & \begin{tabular}[c]{@{}c@{}}HSI
\end{tabular} & 400-2500      & 330  & 30
                            \\
  {GF-2}    &\begin{tabular}[c]{@{}c@{}}MSI
                        \end{tabular} & 450-900      & 4    & 4
                            \\
  { WorldView-2}    &\begin{tabular}[c]{@{}c@{}}MSI
                        \end{tabular} & 450-900     & 4    & 2
                            \\
  {  Hyperion}    &\begin{tabular}[c]{@{}c@{}}HSI
                        \end{tabular} & 400-2500     & 220    & 4
                            \\  \bottomrule
\end{tabular}
\end{center}
\end{table*}

Recently, a number of  approaches have been presented to fuse the HSI and MSI.
However, it lacks critical discussions for newly proposed HSI-MSI fusion methods.  Vivone \emph{et al.} \cite{pansharpeningreview} give a review and comparison for the pan-sharpening approach, and pan-sharpening is a  special instance of the HSI-MSI fusion. This reference mainly focuses on the traditional component substitution (CS) methods and multi-resolution analysis (MRA) methods.
Later, Loncan \emph{et al.} \cite{hyperpsharpening}  present a
study of the HSI-MSI fusion methods, which compares the  newly  proposed matrix factorization (MF) based fusion methods  with
 traditional pan-sharpening approaches.
 More recently, Yokyao \emph{et al.} \cite{Yokoyafusion} give the comparative  study for  more newly proposed  HSI-MSI fusion methods, and select ten representative works for comparison.  However, the recently proposed tensor based fusion approaches and deep convolution neural network (CNN) based approaches are not considered.\par

In this article,  we give a comprehensive review for the HSI-MSI fusion approaches.
 In specific,  HSI-MSI fusion methods are categorized as
 four classes, including methods extended by pan-sharpening, MF based methods, tensor representation (TR) based methods, and deep CNN based methods. Furthermore, we give a detailed introduction to   these methods and  clarify their characteristics, advantages, and limitations. What is more, we also introduce the challenges and guidelines for HSI-MSI fusion.  Compared with the previous works, we give the more detailed  introduction to the recent progress in HSI-MSI fusion, especially on the tensor based fusion methods and deep CNN based approaches.\par

The remainder of this paper is organized  as follows. Section II introduces the notations. In Section III, we review the representative HSI super-resolution literature. Section VI introduces the experimental comparison for the representative fusion methods,
Section V introduces the challenges and possible research direction for HSI-MSI fusion.  Section VI concludes the paper.\par

\section{ Notations}
Before introducing different HSI-MSI fusion methods, we firstly give  notations for the HSI and MSI.
The high-resolution HSI,  low-resolution HSI, and  high-resolution MSI are represented by tensors, and   are denoted by $\mathcal{X}\in \mathbb{R}^{W\times H\times S}$, $\mathcal{Y}\in \mathbb{R}^{w\times h\times S}$, and $\mathcal{Z}\in \mathbb{R}^{W\times H \times s}$, respectively. The first and second modes of tensor index the spatial dimension, and the third mode of the tensor indexes spectral dimension.   ${\bf M}_{(n)}\in \mathbb{R}^{I_n\times I_1I_2,...,I_{n-1} I_{n+1},..., I_N}$ is the matrix obtained by unfolding the tensor $N$-dimensional tensor  $\mathcal{M}\in \mathbb{R}^{I_1\times I_2,...,\times I_N}$ along $n$-th mode. \par

\textbf {Tensor  product}: The $n$-mode product of tensor $\mathcal{M}\in \mathbb{R}^{J_1\times J_2...\times J_N}$ and   matrix ${\bf B}\in \mathbb{R}^{I_n\times J_n}$   is written  as
\begin{equation}\label{eq111}
\mathcal{P}=\mathcal{M}\times{_n}{\bf B}
\end{equation}
 where  $\mathcal{P}\in{\mathbb{R}^{J_1\times J_2...\times I_n...\times J_N}}$, and
 it is equivalent  to the following equation
 \begin{equation}\label{eqmatrix1}
  {\bf P}_{(n)}={\bf B}{\bf M}_{(n)}
    \end{equation}

\textbf {Tucker  Decomposition}:    Based on the Tucker decomposition \cite{Tucker1996}, tensor $\mathcal{M}\in \mathbb{R}^{J_1\times J_2...\times J_N}$ can be factored as
\begin{equation}\label{eq200}
\mathcal{M}=\mathcal{N}\times{_1}{\bf D}_1\times{_2}{\bf D}_2...\times{_N}{\bf D}_N.\\
\end{equation}
 in which ${\bf D}_n\in \mathbb{R}^{J_n\times I_n} (n=1,2,...,N)$ denotes factor matrix in $n$-th mode, and  $\mathcal{N}\in \mathbb{R}^{I_1\times I_2...\times I_N}$ is the core tensor.
\par

\section{HSI-MSI  Fusion Methods}
 According to the  characteristics of HSI-MSI fusion approaches, they can be classified into four categories, that is, pan-sharpening based approaches, MF based approaches, TR based approaches, and deep CNN based approaches. The approaches in each family are introduced in detail in the following context.

 \subsection{Pan-sharpening based HSI-MSI Fusion  Approaches}
  The early spatial-spectral fusion methods aim at combining a low-resolution MSI with a high-resolution panchromatic (PAN) image \cite{deng2019fusion,ma2020pan}, which is referred to as  pan-sharpening. The pan-sharping has a very important and broad application in remote sensing.
  There are two representative types of pan-sharpening approaches, including CS and MRA. Literature \cite{pansharpeningreview} gives a comprehensive review of pan-sharpening methods.
  The CS  approaches firstly up-sample the low-resolution MSI as the same spatial size as that of PAN image, and then separate the spatial information and spectral information of  up-sampled MSI in distinct components based on specific transformation.  Subsequently, the spatial information is replaced by the  PAN image, and fused MSI is obtained by  bringing the replaced spatial information and spectral information back
to the image domain via the inverse transformation. To reduce the distortion brought by this fusion method, histogram matching of the PAN image to the
 corresponding component is implemented  before the replacement. Therefore, the histogram-matched PAN has the same
 variance and mean as the module to be replaced. The representative works in this family consist of intensity hue saturation (IHS) \cite{ihs},  principal component analysis (PCA) \cite{pca}, and Gram-Schmidt (GS) \cite{gs}, which differ by the transformations taken in the fusion procedure. To further enhance the performance of CS approaches,  its adaptive version, called adaptive CS \cite{adpativecs}, has been come up with. In general, the CS approach can be written as
  \begin{equation}
                  \bar{\bf M}_{n}=\hat{\bf M}_{n}+g_n({\bf P}-{\bf M}_w), ~ n=1,...,N
\end{equation}
in which $ \bar{\bf M}_{n}$ and $\hat{\bf M}_{n}$ denote $n$-th band of fused MSI and up-sampled MSI, and $\bf P$ denotes the PAN image, $g_n$ is the injection gains for $n$th band. ${\bf M}_w$ is the weighted average of different spectral bands defined as
   \begin{equation}
                  {\bf M}_{w}=\sum_{n=1}^Nw_n\hat{\bf M}_{n}, ~n=1,...,N
\end{equation}
where $w_n$ is the weight for $n$-th band.  In general, CS approaches are easy and can be implemented efficiently. However, they may  cause significant spectral distortions because of local dissimilarities between the MSI and PAN image. \par

 The MRA methods obtain the fused  MSI by  injecting the high-resolution structures of the PAN image, which is acquired by a
multi-resolution decomposition, into the low-resolution MSI.   The approaches in this family differ from using various multi-resolution decomposition method to extract high-resolution spatial structures from the PAN image, such as,  wavelet transform \cite{wavelet, wavelet1},  Laplacian pyramid (LP) \cite{pyramid},  contourlet \cite{contourlet},  curvelet \cite{curvelet}, and so on. Based on the MRA, the fused MSI can be written as
 \begin{equation}
                  \bar{\bf M}_{n}=\hat{\bf M}_{n}+g_n({\bf P}-{\bf P}_L), ~n=1,...,N
\end{equation}
where ${\bf P}_L$ is the low-pass version of the PAN image. The representative MRA approaches comprise of  low-pass filtering \cite{hpm,hpf} and pyramidal decompositions \cite{mtfglp}.\par

 Since pan-sharpening is a particular instance of the HSI-MSI fusion,
many
trials have been made to extend the  pan-sharpening approaches for fusing HSI and MSI. To adapt the pan-sharpening method to HSI-MSI fusion,
Chen \emph{et al.} \cite{hsipan} firstly divide the spectral bands of HSI into a few groups based on the spectral coverage,  and  then fuse each band of MSI  with the   corresponding spectral bands in  HSI  by
making using of  the existing pan-sharpening approaches. Furthermore,  Selva \emph{et al.} \cite{hypersharpening}  synthesize
the  image of high spatial resolution for every spectral band of HSI   via linear regression on high-resolution MSI, and fuse every spectral band of the HSI with the synthesized high-resolution image    via the pan-sharping method.
  Fusion results verify that the synthesized high-resolution image can obtain much better fusion
results than a selected band in MSI for the fusion. The pan-sharpening based HSI-MSI fusion methods   often have low computation cost and can be implemented fast. However, they often produce remarkable distortions when the spatial resolutions of  HSI and  MSI differ greatly.

\begin{table*}[t]\small
\caption {The properties of the representative MF based fusion methods}
\vspace{-1mm}
\begin{center}
\begin{tabular}{l c c c c  c}
 \toprule
{Method}

 &Spectral prior   &Spectral basis    &Optimization category   \\
 \midrule

SMF  \cite{mff} &Sparsity   & $\ell_1$ minimization    &1                 \\
SASFM  \cite{ssfm} &Sparsity   & K-SVD    &1                 \\
 GSOMP  \cite{gsomp}     &Sparsity   &Online dictionary learning     &1                 \\
 BSR \cite{Akhtar2015Bayesian}        &Sparsity    &Online dictionary learning      &1       \\

Fuse-S   \cite{fuses}    &Low rank   &PCA    &2      \\
  NSSR   \cite{nssr}   &Sparsity   &Non-negative dictionary learning     &2       \\
  Hysure   \cite{hysure}   &Low rank   &SVD/VCA     &2        \\
   CNMF   \cite{cnmf}   &Low rank    &VCA     &3        \\
CSU \cite{csu}   &Low rank    &SISAL   &3    \\
   SSSR   \cite{dian2019multispectral}   &Sparsity   &Non-negative dictionary learning     &3       \\
    FUMI   \cite{fumi}   &Low rank   &VCA     &3       \\
  \bottomrule
 \end{tabular}
  \end{center}
 \label{tab_MF}
\end{table*}
 \subsection{MF based HSI-MSI Fusion Approaches}
 The MF based HSI-MSI fusion approaches unfold the three-dimensional fused HSI  ${\cal X}$ with the spectral mode, and obtain the matrix ${\bf X}_{(3)}\in \mathbb{R}^{S\times WH}$. The approach in this category assume that  ${\bf X}_{(3)}$ can be decomposed as spectral basis  ${\bf D}\in \mathbb{R}^{S\times L}$  multiplied by coefficients ${\bf A}\in \mathbb{R}^{L\times WH}$, denoted by
                  \begin{equation}
                  {\bf X}_{(3)}={\bf D}{\bf A},
\end{equation}
 Both the high-resolution MSI  and   low-resolution HSI can be regarded as the subsampled versions of  the fused HSI:
 \begin{equation}\label{eqdfd}
 \begin{split}
 {\bf Y}_{(3)}={\bf X}_{(3)}{\bf G},\\
 {\bf Z}_{(3)}={\bf P}_{3}{\bf X}_{(3)},
 \end{split}
 \end{equation}
 where   ${\bf Z}_{(3)}\in \mathbb{R}^{s\times WH}$ and ${\bf Y}_{(3)}\in \mathbb{R}^{S\times wh}$ are matrices acquired by    unfolding   $\mathcal{Z}$ and  $\mathcal{Y}$ along the third mode, respectively. Matrix ${\bf G} $ models the spatial degradation procedure.
  Based on the above formulation, the fusion of HSI and MSI is converted to the computation of   coefficients and spectral basis.

In general,  spectral basis $\bf D$ denotes the spectral
information of the high-resolution HSI. On the basis of the
way to model the spectral basis, the  MF based approaches often
 can be classified as sparse representation \cite{gsomp, ssfm, nssr}
 methods and low-rank methods \cite{hysure,fuses,cnmf}.
 The sparse representation methods regard the spectral
 basis ${\bf D}\in \mathbb{R}^{S\times L}$ as the over-complete
 dictionary, and the number of atoms $L$ is often larger than
  $S$ in order to obtain the sparsity. They assume that each spectral
   signature is  a linear combination of a few atoms in the dictionary.
   The dictionaries are often learned from the low-resolution HSI via sparse
   dictionary learning algorithms, such as K-SVD \cite{ksvd}, online dictionary
   learning \cite{Mairal2009Online}, and non-negative learning \cite{nssr}.
   Then the estimation of coefficients is regularized by sparse prior, and
   they are often estimated by a sparse coding algorithm. The low-rank based methods
   consider that  spectral signatures can be represented by a  low-dimensional
   subspace, and $\bf D$ is a low-rank matrix with $L<<S$. The low-rank spectral
    basis $\bf D$  is often learned from the low-resolution via
    vertex component analysis (VCA)   \cite{nascimento2005vertex}, simplex identification
via split augmented
Lagrangian (SISAL) \cite{sisal}, principal components analysis, or
    truncated singular value decomposition (SVD). In essence, both sparse
    representation and low-rank representation based methods are on purpose
     of modeling the similarities and redundancies of among the spectral bands,
      and both of them can well preserve the spectral properties. However,
       the low-rank representation based methods can largely
       reduce the dimension of spectral mode, and can achieve much faster
       fusion compared with the sparse representation approaches.

The MF based fusion methods intend to estimate the spectral basis and coefficients  by solving the corresponding optimization problem.  Based on the optimization formulation for the spectral basis and coefficients, the MF based fusion methods mainly has three classes.   The approaches in the first  family  argue that the spectral information and spatial information mainly relies on the low-resolution HSI and high-resolution MSI, respectively. Based on this assumption, they
 estimate the spectral basis only from the observed HSI, and then obtain the coefficients only from the observed MSI. The optimization formulation can be written as
 \begin{equation}\label{method1}
 \begin{split}
 \min_{{\bf D}}||{\bf Y}_{(3)}-{\bf DA}{\bf B}{\bf S}||_F^2+\lambda_1\psi({\bf D}),\\
\min_{{\bf A}}
||{\bf Z}_{(3)}-{\bf R}{\bf DA}||_F^2+\lambda_2\phi({\bf A}),
 \end{split}
 \end{equation}
 in which $\lambda_1\psi({\bf D})$ and  $\lambda_2\phi({\bf A})$ are regularization term on $\bf D$ and $\bf A$, respectively. Most of the early works in MF belong to the first category.
 For example, reference \cite{mff} firstly introduces  the sparse MF (SMF) for HSI-MSI fusion, where they estimate spectral dictionary with sparse dictionary learning method, and estimate the sparse coefficients by sparse coding algorithm on the MSI.
   Huang \emph{et al.}  \cite{ssfm} propose a similar idea, called as the SASFM, in which the spectral basis is learned via  K-SVD \cite{ksvd}.
  Akhtar \emph{et al.} \cite{gsomp} make use of the local
  similarities of the fused HSI, and design a  simultaneous
  greedy pursuit algorithm to compute the sparse coefficients.
      The spectral dictionary is learned  with a Beta process from the
       low-resolution HSI  \cite{Akhtar2015Bayesian},
       and coefficients are obtained by Bayesian sparse coding
        on the high-resolution MSI.   \par

 The methods in the second category argue that low-resolution HSI also contains the spatial information and can contribute the estimation coefficients $\bf A$. They often firstly compute the spectral basis from observed  HSI,
 and then calculate  coefficients from both two images. Based on the maximum a posteriori (MAP),  we formulate the calculation of $\bf A$ ¡¡ as   minimizing a
 function made up of   a regularization term and two quadratic data-fitting terms,
 \begin{equation}\label{eq4333}
 \begin{split}
 \min_{{\bf A}}||{\bf Y}_{(3)}-{\bf DA}{\bf B}{\bf S}||_F^2+
||{\bf Z}_{(3)}-{\bf R}{\bf DA}||_F^2+\lambda_2\phi({\bf A}),
 \end{split}
 \end{equation}
 where $\lambda_2\phi({\bf A})$ denotes the regularization term on the coefficients and models the prior information. Methods proposed in \cite{hysure,fuses} make use of the low-rank prior of HSI along spectral dimensional, and regard $\bf D$ as the low-dimensional subspace, which is obtained from the observed  HSI by vertex component analysis   \cite{nascimento2005vertex} or truncated singular value decomposition. For example, Simo\~es \emph{et al.} \cite{hysure} exploit the total variation as the regularizer, which promote the spatial smoothness.  Wei \cite{fuses} present a patch-based sparse prior to preserve the self-similarities.
 Dong \emph{et al.}  \cite{nssr} firstly learn the over-complete dictionary from the low-resolution HSI, and then use the prior of non-local similarities and sparse prior to regularize the estimation of coefficients. To model the global similarities of the high-resolution HSI,  Han \emph{et al.} \cite{hanself} group the similar patches, and assume that the given patch can be linearly represented by the similar patches. Besides, they also segment the high-resolution HSI based on the super-pixel to learn the local similarities of the high-resolution HSI. Zhou \emph{et al.}  \cite{zhouyuan} and Veganzones \emph{et al.}  \cite{Veganzones2016} use the local low-rank regularization to learn local similarities of the fused HSI, and reconstruct each local region independently. To compute the coefficients fast, Wei \emph{et al.}  \cite{fusewei} formulate the fusion problem as tackling a Sylvester Equation, which has a closed solution and gets rid of iteration.

 The   methods in  third  category solve the fusion problem based on the coupled matrix decomposition, which is not based on the fixed dictionary, and alteratively update the spectral basis and coefficients. The representative work is coupled nonnegative MF (CNMF) \cite{cnmf}. Based on non-negative MF \cite{nmf},  the CNMF  alteratively factors the low-resolution HSI and high-resolution MSI as follows,
  \begin{equation}\label{eq43}
 \begin{split}
 {\bf Y}_{(3)}={\bf D}{\bf A}_m,\\
 {\bf Z}_{(3)}={\bf D}_m{\bf A},
 \end{split}
 \end{equation}
 where the factor matrices ${\bf A}_m$ and   ${\bf D}_m$  are initialized  as ${\bf A}_m ={\bf ABS}$ and ${\bf D}_m ={\bf RD}$  in each factorization of ${\bf Y}_{(3)}$ and ${\bf Z}_{(3)}$. By exploiting a number of the  priors in spectral unmixing,  Lanaras \emph{et al.}  \cite{Lanaras} estimate ${\bf D}$ and ${\bf A}$ via the  proximal alternating linearized minimisation. To obtain more accurate estimation of spectral basis and coefficients, some approaches calculate them from both two observed  images, and formulate the fusion problem as
\begin{equation}\label{eqsly33}
 \begin{split}
 \min_{{\bf D},{\bf A}}||{\bf Y}_{(3)}-{\bf DA}{\bf B}{\bf S}||_F^2+
||{\bf Z}_{(3)}-{\bf R}{\bf DA}||_F^2+\lambda_1\psi({\bf D})+\lambda_2\phi({\bf A}),
 \end{split}
 \end{equation}
 where $\lambda_1\psi({\bf D})$ and  $\lambda_2\phi({\bf A})$ are prior regularization term on ${\bf D}$ and ${\bf A}$, respectively.
 The optimization problem \eqref{eqsly33} is solved by alternatively optimizing $\bf D$ and  $\bf A$, that is
 \begin{equation}\label{eqsly33}
 \begin{split}
 \min_{{\bf D}}||{\bf Y}_{(3)}-{\bf DA}{\bf B}{\bf S}||_F^2+
||{\bf Z}_{(3)}-{\bf R}{\bf DA}||_F^2+\lambda_1\psi({\bf D}),\\
\min_{{\bf A}}||{\bf Y}_{(3)}-{\bf DA}{\bf B}{\bf S}||_F^2+
||{\bf Z}_{(3)}-{\bf R}{\bf DA}||_F^2+\lambda_2\phi({\bf A}),
 \end{split}
 \end{equation}
 where $\lambda_1\psi({\bf D})$ and  $\lambda_2\phi({\bf A})$ are regularization term on $\bf D$ and $\bf A$, respectively. The two sub-problems are often solved by the alternating direction
method of multipliers  \cite{boyd2011distributed}. The main differences of methods in this category is using different regularization on $\bf D$ and $\bf A$.
   Wycoff \emph{et al.} \cite{snnmf} use $\ell_1$ norm based regularization
   to promote the sparse representation, the spectral basis and coefficients are
    iteratively estimated by alternating direction method of multipliers.
    Wei \emph{et al.} \cite{fumi} introduce the  physical properties in
    spectral unmixing, including non-negative and sum-to-one constraints,
     for the spectral basis and coefficients.
      Dian \emph{et al.} \cite{dian2019multispectral} propose a method, called as SSSR,
      which assumes that
       a pixel can be linearly expressed by similar pixels,
        and uses this assumption and sparse prior to regularize the
         estimation of the coefficients. \par

Table  \ref{tab_MF} summarizes the properties of some representative MF based approaches.
  Although MF based approaches  have obtained  superior performance than the pan-sharpening based methods, they suffer from three disadvantages. First of all, these methods often need to solve the complex optimization problem iteratively, and the computation cost is high. Secondly, the performance of them is usually very sensitive to the parameter selection, and the parameters are hard to set. Finally, they are based on the observation model \eqref{eqdfd}, and therefore
   highly rely on accurate estimation of the point spread function (PSF) and spectral response function (SRF).


\subsection{TR based HSI-MSI Fusion Approaches}
The HSIs and MSIs have three dimensions, and therefore can be expressed by a three-dimensional tensor. Based on this fact, TR has been an active topic for HSI-MSI fusion. The approaches  in this category  are based on different kinds of TR methods.\par

Tucker decomposition \cite{Tucker1996} is one of the  widely used TR methods, and it decomposes the high-dimensional tensor as factor matric of each dimension and a core tensor. Tucker decompositon can separate the  information of each dimension in each tensor, meanwhile it can also establish the correlations among the information of each dimension via the core tensor. Benefiting the mentioned advantage,
 Tucker decomposition has been actively studied   in  completion \cite{wuafused,zhao2016Bayesian}, visual tracking \cite{Chen2017Trifocal,Ma2016Discriminative}, objection detection \cite{Wong2015},    and compressive sensing \cite{feng2016compressive}. Recently, it has also shown promising performance in many restoration tasks \cite{Peng2014Decomposable,xuenonlocallowtran,Yang2015Compressive,xue2019nonlocal}.
 Based on the Tucker decomposition, Dian \emph{et al.} \cite{nlstf,nlstf_smbf} firstly assume that  the  high-resolution HSI  $\mathcal{X}$  can be factored as
\begin{equation}\label{eq204}
\mathcal{X}=\mathcal{C}\times{_1}{\bf W}\times{_2}{\bf H}\times{_3}{\bf S},
\end{equation}
where the matrices ${\bf W}\in \mathbb{R}^{W\times n_w}$, ${\bf H}\in \mathbb{R}^{H\times n_h}$, and ${\bf S}\in \mathbb{R}^{S\times n_s}$ denote the dictionaries of three dimensions.  The dictionaries ${\bf W}$ and ${\bf H}$  express the basic information of two spatial dimensions, and  dictionary ${\bf S}$ expresses the basic information of spectral dimension.
 ${\mathcal{C}\in \mathbb{R}^{n_w\times n_h\times n_s}}$ is the core tensor in the Tucker decomposition, which depicts the relationships of information of three dimensions. The PSF of HSI imaging sensor is often Gaussian kernel, which can be decomposed in two spatial modes. Based on this assumption, the low-resolution HSI is expressed as
\begin{equation}\label{eq44}
\mathcal{Y}=\mathcal{X}\times{_1}{\bf S_1}\times{_2}{\bf S_2},
\end{equation}
where ${\bf S_1}\in \mathbb{R}^{w\times W}$ and ${\bf S_2}\in \mathbb{R}^{h\times H}$ denote the subsampling operation on the width and height modes, respectively, and satisfy ${\bf G}=({\bf S_2}\otimes{\bf S_1})^T$.
The high-resolution MSI $\mathcal{Z}$  is given by,
\begin{equation}\label{eq10}
\mathcal{Z}=\mathcal{X}\times{_3}{\bf P_3},
\end{equation}
where ${\bf P}_{3} \in \mathbb{R}^{s\times S}$ is the spectral subsampling  matrix of the multispectral imaging sensor.
   According to equation (\ref{eq204}), the fusion task is calculating core tensor $\cal C$, and  dictionaries $\bf W$, $\bf H$, $\bf S$. Reference \cite{nlstf} firstly partion the high-resolution HSI as many cubes, and group these small cubes based on the leaned structure in high-resolution MSI. They assume that cubes in the same group may be sparsely expressed by  the same dictionaries. For each cube, the dictionaries $\bf W$ and $\bf H$ are learned from high-resolution MSI, and dictionaries $\bf S$ is estimated  from low-resolution HSI. The core tensor is learned by inducing a spare prior. Furthermore, Li \emph{et al.} \cite{cstf} present a coupled sparse Tucker decomposition (CSTF) scheme for HSI-MSI fusion, which estimates the core tensor and dictionary of each mode via proximal altercating optimization. To reduce the computational cost,
   Pr¨¦vost \emph{et al.} \cite{SCOTT} make use of the  truncated SVD to obtain the dictionaries of three modes and estimate the core tensor via solving the  generalized Sylvester equation.
     Chang \emph{et al.}  \cite{wlrtr} also group similar cubes together, and impose a weighted sparsity constraint on the core tensor of similar cubes to make use of the non-local similarities.\par

     Canonical polyadic (CP) decomposition is regarded as a special instance of Tucker decomposition, which can decompose a $N$-dimension tensor as $N$ factor matrices.
      Based on the CP decomposition, Kanatsoulis \emph{et al.} \cite{ctf}  factor the high-resolution HSI as
      \begin{equation}\label{eqcp}
\mathcal{X}=\sum_{i=1}^I{\bf A}(:,i)\circ {\bf B}(:,i) \circ {\bf C}(:,i),
\end{equation}
in which $\circ$ denotes out product of the vector, and
 ${\bf A}(:,i)$ represent $i$-th column of  factor matric ${\bf A}$.   They estimate the three factor matrices  by tackling a least squares problem. In addition, they also
 investigate the semi-blind fusion case, in which the spatial sub-sampling process is not assumed known.     Xu  \emph{et al.} \cite{nctcp} argue that the observed  HSI and  MSI share the same factor matrices in  CP decompositon, and propose a  non-local  CP decomposition for HSI-MSI fusion. \par

   In addition to Tucker decomposition and CP decomposition, many other TR methods have also been actively studied, including tensor singular value decomposition (t-SVD) \cite{tsvd} and tensor-train decomposition \cite{tensortrain}. The t-SVD defines a novel tensor-tensor product, which is on the basis of the vector circular convolution.   According to the  tensor-tensor product,  Xu \emph{et al.} \cite{tpsrsf} first group non-local similar patches to form a three-dimensional tensor $\mathcal{X}_{\{k\}}$, and factor the tensor $\mathcal{X}_{\{k\}}$ as
   \begin{equation}\label{eqtt}
\mathcal{X}_{\{k\}}=\mathcal{D}_{\{k\}}*{\cal{B}}_{\{k\}},
\end{equation}
where * represents the tensor-tensor product  defined in \cite{tsvd}, and $\mathcal{D}_{\{k\}}$ and $\mathcal{B}_{\{k\}}$ represents the tensor dictionary and tensor coefficients.
Based on the above formulation, HSI-MSI fusion be equivalent to the calculation of $\mathcal{D}_k$ and $\mathcal{B}_K$, and the authors impose a sparse prior on the coefficients to preserve the non-local similarities. On the basis of  t-SVD, a new tensor multi-rank is come up with to measure structural complexity and informational of a tensor.
 Reference \cite{ltmr} combines the subspace representation and low tensor multi-rank prior to fuse HSI and MSI. The low-dimension subspace representation factors the  fused HSI as   coefficients and a low-dimensional spectral subspace,  which exploits the similarities in spectral dimension, and  largely decrease the dimension of spectral mode. The spectral subspace is calculated via SVD from the observed  HSI, and the coefficients are computed by  low multi-rank regularization.
Tensor-train decomposition is a very popular TR method, and it defines a new tensor rank, called as tensor-train rank, which  is made up of ranks of matrices by folding tensor along permutations of modes. Dian \emph{et al.} \cite{lttr} present a low tensor-train rank regularized HSI-MSI fusion approach. They firstly group similar full-band patches to form a four-dimensional tensor and then give the relaxed low tensor-train rank constraint to the four-dimensional tensor to make use of  the non-local spatial-spectral similarities.\par

The TR approaches have reported  excellent fusion results on the simulated data fusion. Despite this, the computational cost of them is still very high compared with the pan-sharpening based methods. In addition, they also need to estimate the PSF and SRF of the sensors accurately.

\subsection{Deep CNN based HSI-MSI Fusion}
 Recently,  CNN   has received more and more attention in many image processing applications due to its high efficiency and promising  performance. The CNN is data-driven, and can effectively learn  the various image features from the training data.
 Dong \emph{et al.}  \cite{dong2014learning, dong2016image}  firstly propose a  deep CNN, named as SRCNN, for single image super-resolution, which achieves  the superior performance.
 A series of deep CNN based pan-sharpening methods \cite{GiuseppePansharpening,pannet}  are presented  to combine the observed MSI with the high-resolution  panchromatic image to obtain high-resolution MSI. The deep CNN based pan-sharpening methods can be easily used for  HSI-MSI fusion by changing the  number of filters  in the first and the last convolution layers.
 The deep CNN based HSI-MSI fusion approaches intend to learn the following non-linear mapping function $f$,
\begin{equation}\label{eqCNN}
{\cal X}={f}({\cal Y},{\cal Z},\Theta),
\end{equation}
in which $\Theta$ is the parameter of the CNN.
 The deep CNN based fusion approaches can be categorized as one-branch CNN based fusion approaches and  two-branch CNN approaches. The  one-branch CNN based fusion approaches firstly combine the features of  the  high-resolution MSI and low-resolution HSI, and then input them into the one-branch CNN to map the high-resolution HSI. For example, the PanNet \cite{pannet} firstly concatenates the features of   up-sampled observed HSI with observed MSI in high-pass domain. In addition,
 instead of directly mapping the observed HSI and observed MSI to the fused HSI, Dian \emph{et al.} \cite{DHSIS}  initialize the fused HSI based on the observational models, and then map the initialized high-resolution HSI to desired high-resolution HSI via the deep CNN.  Xie \emph{et al.} \cite{mhfnet} combine  the low-rank constraint  of the high-resolution HSI,  observation models of the   MSI and HSI, and image prior learned via deep CNN, and formulate the fusion task as a new optimization problem.
 Based on the proximal gradient method, they design the deep CNN to solve the optimization problem in an iterative way. Xie \emph{et al.} \cite{hpdp} firstly upsample the low-resolution HSI by using a CNN, and then train the fusion CNN in the high-pass domain. Finally, they combine the output of fusion CNN with the observation by tacking a Sylvester equation. The CNN based fusion methods are data-driven, often need sufficient HSI and MSI data pairs for the training procedure. However, the training data are often not available. Besides, the  generalization ability of these methods is often limited by the imaging models and types of training data.
  To solve the above problem,  reference \cite{cnnfus} introduces a CNN denoiser based HSI-MSI fusion method, which uses the CNN for denoising trained on the grayscaling image. Besides, this method also combines the observation model of the HSI and MSI into consideration, and it can flexibly deal with different types of data.
 \par

 The  two-branch CNN based fusion approaches use two sub-networks to extract features from low-resolution HSI and high-resolution MSI, respectively, and then combine the two features to predict the high-resolution HSI.
 Yang \emph{et al.} \cite{yang2018hyperspectral}  propose  a CNN based fusion method with two paths, which exploits the  two paths  to acquire  the spatial features and spectral features   from the high-resolution HSI and  low-resolution MSI, respectively.
 Wang  \emph{et al.} \cite{DBIN} design a
 iterative refinement unit to iteratively make use of the observed HSI and  MSI to refine the fused HSI. Considering that the spatial structures of the observed HSI and  MSI are very different,  Han  \emph{et al.} \cite{ssfusion} introduce a multi-scale CNN method, and it contains  the spatial  reservation pathway and spectral reservation pathway, which
   gradually reduces and increases  the feature sizes of the   low-resolution HSI  and high-resolution MSI, respectively.\par

There are three  advantages of deep CNN based HSI-MSI fusion methods. Firstly, they often do  not need to know the PSF and SRF of the tensor, which may be hard to know in practice. Secondly, compared with the MF and TR based methods, deep CNN based fusion method can be implemented much faster, since they do not need iteration and can be easily  accelerated via high-performance GPU. What is more, the non-linear function and convolution layer can make CNN   ¡¡has a powerful ability to  learn the image features and achieve robust fusion.
   Although the deep CNN based approaches have achieved promising results and high speed for HSI-MSI fusion. However, they still have two disadvantages. Firstly, methods in this family need additional data  for training. However, it often lacks the available low-resolution HSI, high-resolution-MSI, and high-resolution HSI for pre-training. The common way to produce the training data is regarding the available HSI as a reference image, and down-sample it to produce the low-resolution HSI and high-resolution MSI, which may not conform to the practical condition. What is more, the generalizability of the deep CNN is also a big  challenge. Since the number of spectral bands, spatial resolution, and  spectral coverage of the data may be different, CNN trained on one kind of data  can  not be applied to the other kinds of data.

\begin{table*}[t]\small
\caption {The properties of compared approaches}
\vspace{-1mm}
\begin{center}
\begin{tabular}{l c c c c c }
 \toprule
{Method}

  &Category  &Pre-training  &PSF  &SRF   \\
 \midrule

 GSA  \cite{adpativecs} &CS    &No     &No                 &No\\
 GLP-HS \cite{mtfglp}     &MRA   &No     &No    &No              \\
 NSSR \cite{nssr}    &MF    &No     &Yes    &Yes   \\
  CNMF   \cite{cnmf}   &MF    &No     &Yes    &Yes    \\
CSU \cite{csu}   &MF    &No     &Yes  &Yes  \\
Fuse-S   \cite{fuses}    &MF   &No     &Yes   &Yes    \\
 CSTF   \cite{cstf}   &TR    &No     &Yes  &Yes  \\
  LTMR   \cite{ltmr}   &TR   &No     &Yes    &Yes    \\
  CNN-Fus \cite{cnnfus}&CNN   &Yes &Yes  &Yes  \\
  \bottomrule
 \end{tabular}
  \end{center}
 \label{tab_methods}
\end{table*}

   \section{Experiments}
\subsection{Testing Approaches}
This paper uses nine representative methods for comparison. Among these methods, consisting of  Gram-Schmidt adaptive (GSA) \cite{adpativecs}, generalized Laplacian pyramid (GLP-HS) \cite{mtfglp}, non-negative structured sparse representation (NSSR) \cite{nssr}, CNMF \cite{cnmf}, coupled spectral unmixing (CSU) \cite{csu}, sparse representation (Fuse-S) \cite{fuses},
CSTF, low tensor multi-rank based method (LTMR) \cite{ltmr}, and CNN fusion method (CNN-Fus) \cite{cnnfus}.
Among these methods, we use Gram-Schmidt adaptive (GSA) as a representative CS-based fusion method, and the generalized Laplacian pyramid is the MRA-based method.
NSSR, CNMF, CSU, and Fuse-S belong to the category of MF, where NSSR is on the basis of spectral sparse representation, and the CNMF, CSU, and Fuse-S are based on low-rank representation. CSTF and LTMR  belong to the category of TR, in which CSTF is on the basis of sparse Tucker decomposition, and LTMR is based on low tensor multi-rank representation. CNN-Fus belongs to the category of deep CNN. The properties of the testing methods are shown in Table \ref{tab_methods}.

   \subsection{Quality Metrics}
   For the HSI-MSI fusion  approaches, the ideal high-resolution HSI is unknown, which makes it hard to evaluate the quality of the fused image directly. To tackle this problem, two assessment procedures have been proposed. The first one considers an available HSI as the ideal fused HSI and  simulates the high-resolution MSI and low-resolution HSI  via  spectrally downsampling and spatially downsampling, respectively.
   In this way, the HSI-MSI fusion approaches can be applied to the two simulated data to obtain the fused HSI. Therefore, the quality of the fused HSI can be evaluated via the quality metrics based on the available reference image. The choice of the downsampling filter is very important for the simulation. The spatial down-sampling filter  should match the  PSF of the HSI imaging sensor. The spatial degradation procedure is often simulated by firstly  applying a Gaussian filter with zero mean and then  conducting the uniform sub-sampling. The spectral down-sampling filter should match the spectral response function of the MSI imaging sensor and  is represented by the matrix as in \eqref{eqdfd}.
Although the above operation can obtain
accurate evaluation of the fused image, the mismatches of performance  between the simulated data fusion and real fusion may  exist.
 Therefore, an evaluation  via visual
inspection is also a very crucial step for  identifying the spatial dissertations and
spectral distortions in the fused images.

%

Before introducing the quality metrics, we first give some notions. Let $\cal X$ and $\hat{\cal X}$ denote the reference HSI and fused HSI, respectively. ${\cal X}_i$ and ${\cal X}^i$ denote $i$-th band and $i$-th spectral pixel of the $\cal X$, respectively.
Given the reference image ${\cal X}$, seven quality indexes are unusually used to evaluate the quality of the fused HSI $\hat{\cal X}$. All elements of the image are scaled to [0, 255] when calculating the quality metrics.


\subsubsection{PSNR}
The peak signal to noise ratio (PSNR) is a very popular quality metric, for two grayscaling images $\bf A$ and $\bf B$
\begin{eqnarray}
 \text{PSNR}({\bf A},\bf B)=20\text{log}\frac{255}{\text{RMSE}({\bf A},{\bf B})},
\end{eqnarray}
in which the root mean square error (RMSE)  is expressed  as
\begin{eqnarray}
 \text{RMSE}({\bf A},\bf B)=\sqrt{\frac {||{\bf A}-{\bf B}||_F^2}{N}},
\end{eqnarray}
in which $N$ represents the number of pixels. The PSNR for HSI is defined as average value of all bands.
The bigger value of PSNR means better fusion results.

\subsubsection{ERGAS}
The relative dimensionless global error in synthesis (ERGAS) \cite{Wald}  is represented as
\begin{eqnarray}
\text{ERGAS}(\mathcal{X},\widehat{\mathcal{X}})=\frac{100}{d}\sqrt {\frac{1}{S}\sum_{i=1}^S\Big(\frac{\text{RMSE}(\mathcal{X}_i,\hat{\cal X}_i)}{{\mu}({\hat{\cal X}_i})}\Big)^2},
\end{eqnarray}
in which $d$ is spatially sub-sampling factor,   and ${\mu}$ represents the mean value of the image.  The smaller ERGAS, the better
the fusion results.

\subsubsection{SAM}
The SAM is a very crucial index to evaluate the spectral distortions, which is defined as
\begin{eqnarray}\label{eq3000}
\text{SAM}(\mathcal{X},\hat{\mathcal{X}})=\frac{1}{M}\sum_{j=1}^{M} \text{arcos} \frac{{\hat{\cal X}^j}\cdot{\cal X}^j}{||{\cal X}^j||_2||\hat{\cal X}^j||_2}.
\end{eqnarray}
in which $M$ is the number of spectral pixels, and $\cdot$ denotes inner product of two vectors.
A smaller value of SAM means fewer spectral distortions.

\subsubsection{UIQI}
To overcome the some disadvantages of RMSE, Wang \emph{et al.} \cite{Wang2002} prose a index called as Universal Image Quality Index (UIQI) or Q index. The UIQI between two images is calculated as the average value of all image patches, and
 the UIQI for two image patches $\bf a$ and $\bf b$ is defined as
\begin{eqnarray}
\text{Q}({\bf a},{\bf b})=  \frac{4\mu_{{\bf a}}\mu_{{\bf b}}}{\mu_{\bf a}^2+\mu_{\bf b}^2}\frac{\sigma^2_{{\bf a},{\bf b}}}{\sigma_{\bf a}^2+\sigma_{\bf b}^2},
\end{eqnarray}
in which   $\sigma$ and $\mu$ represent  the variance and mean, respectively.
The average value of all spectral bands is the UIQI for HSI.
The larger value of UIQI means better fusion results.
\par


\subsubsection{SSIM} The  structural similarity index (SSIM) \cite{ssim} is a widely used image quality metric, which  measures the structural similarities between the estimated image and the reference image. The SSIM for HSI  is computed in each band, and then obtains the average value of all bands.

\subsubsection{T}
Since the high dimension of the HSI, the computation efficiency is also a very crucial index for the HSI-MSI fusion. It can be quantified by the time complexities and the computational time when the methods are implemented. In this paper, we calculate the running time in seconds of the compared approaches, and this index is denoted as $T$. To ensure a fair comparison, all approaches are coded at Matlab 2018  computer equipped with  8-GB random access memory and  Intel Core-i5-9300H CPU with 2.4-GHz.   \par

\begin{table*}[t]
\caption {Quantitative indexes of the testing methods  on Cuprite Mine \cite{561}}
\vspace{-1mm}
\begin{center}
\begin{tabular}{l c c c c c c c }
 \toprule
 \multirow{2}{*}{Method}
 &
 \multicolumn{6}{c}{{Cuprite Mine} }
  \\
 \cline{2-7}
    &PSNR  &ERGAS &SAM &UIQI  &SSIM  &T \\
 \midrule
 Best Values  &$+\infty$ &0 &0 &1 &1  &0\\
 GSA  \cite{adpativecs} &38.449    &2.000    &1.838    &0.860    &0.933                    &  \textbf{6.303}\\
 GLP-HS \cite{mtfglp}     &35.372    &2.319    &2.038    &0.793    &0.901       &40.951            \\
 NSSR \cite{nssr}    &36.532    &2.435    &2.141    &0.806    &0.890   &486.428\\
  CNMF   \cite{cnmf}   &42.329    &2.253    &1.267    &0.926    &0.959    &217.413\\
CSU \cite{csu}   &42.171    &2.353    &1.299   &0.927   &0.958    &625.130\\
Fuse-S   \cite{fuses}    &43.714    &1.869    &1.150    &0.942    &0.9680    &852.530\\
 CSTF   \cite{cstf}   &42.733    &\textbf{1.693}    &1.179    &0.935   &0.962   &482.204\\
  LTMR   \cite{ltmr}   &42.329    &2.253    &1.267    &0.926    &0.959   &311.006\\
  CNN-Fus \cite{cnnfus}& \textbf{44.223 }   &{ 2.113} &\textbf{ 1.131}  &\textbf{0.946}  &\textbf{0.971 }   &526.782\\
  \bottomrule
 \end{tabular}
  \end{center}
 \label{tabcuprite}
\end{table*}

\begin{table*}[t]\small
\caption {Quantitative indexes of the testing methods  on  Pavia University \cite{561}}
\begin{center}
\begin{tabular}{l c c c c c c c }
 \toprule
 \multirow{2}{*}{Method}
 &
 \multicolumn{6}{c}{{Pavia University} }
  \\
 \cline{2-7}
    &PSNR  &ERGAS &SAM &UIQI  &SSIM  &T  \\
 \midrule
 Best Values  &$+\infty$ &0 &0 &1 &1   &0\\
 GSA  \cite{adpativecs} &35.167    &2.2940    &4.582   &0.967    &0.962                     & \textbf{2.146}\\
 GLP-HS \cite{mtfglp}    &28.559    &4.673    &5.724   &0.884    &0.884      &5.789          \\
 NSSR \cite{nssr}    &  40.329    &1.263    &3.138    &0.982    &0.974    & 242.018\\
  CNMF   \cite{cnmf}   &42.416    &0.997    &2.348   &0.990    &0.985    &120.043\\
CSU \cite{csu}   &  40.492     &1.244     &2.741     &0.986     &0.979    &462.859\\
Fuse-S   \cite{fuses}   &  42.964    &0.941    &2.289    &0.991    &0.985   &660.153\\
 CSTF   \cite{cstf}   &  41.762    &1.081    &2.464    &0.988    &0.980   &438.154\\
  LTMR   \cite{ltmr}   & 42.416    &0.997    &2.348    &0.990    &0.985  & 198.347\\
  CNN-Fus \cite{cnnfus}& \textbf{42.987 }   &\textbf{0.926} &\textbf{ 2.225}  &\textbf{0.992}  &\textbf{ 0.987 }  &105.968\\
  \bottomrule
 \end{tabular}
  \end{center}
 \label{tabpavia}
\end{table*}

\begin{figure*}[!t]
\centering
\subfigure{
\includegraphics[width=29mm]{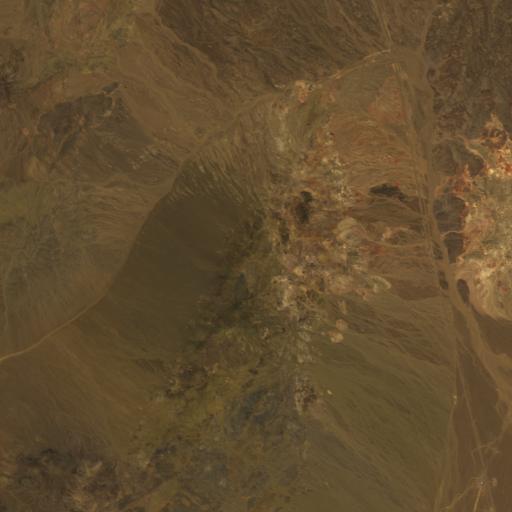}}
\subfigure{
\includegraphics[width=29mm]{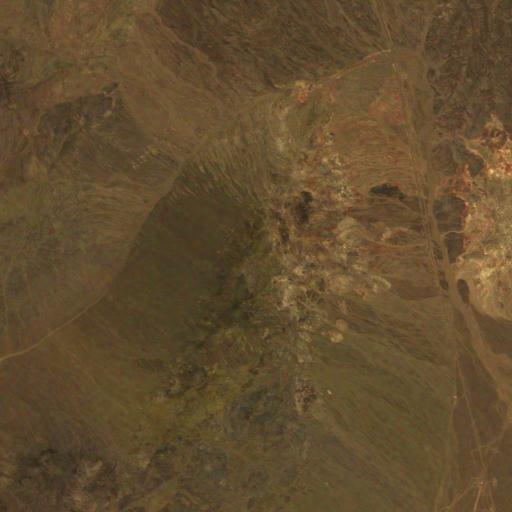}}
\subfigure{
\includegraphics[width=29mm]{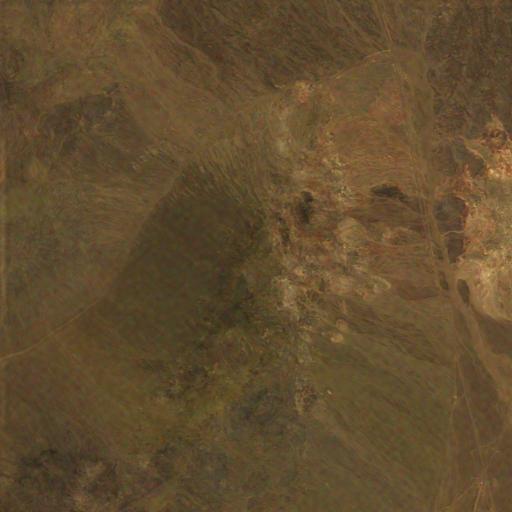}}
\subfigure{
\includegraphics[width=29mm]{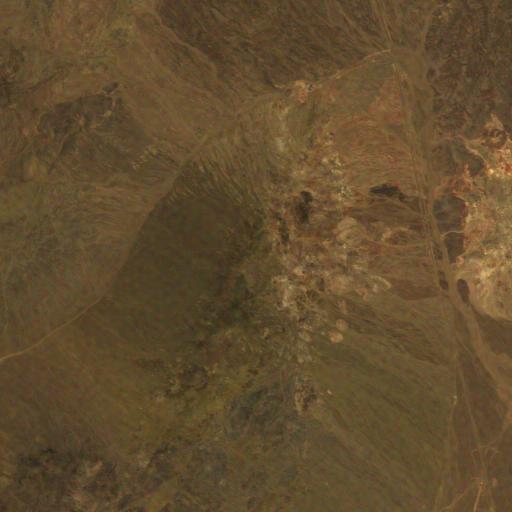}}
\subfigure{
\includegraphics[width=29mm]{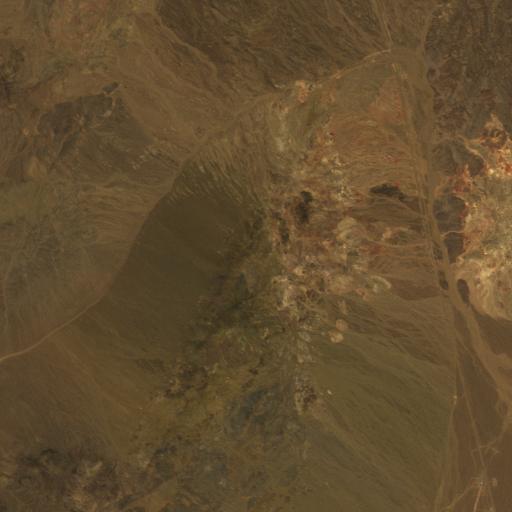}}\\
\setcounter{subfigure}{0}
\subfigure[Reference image]{
\includegraphics[width=29mm]{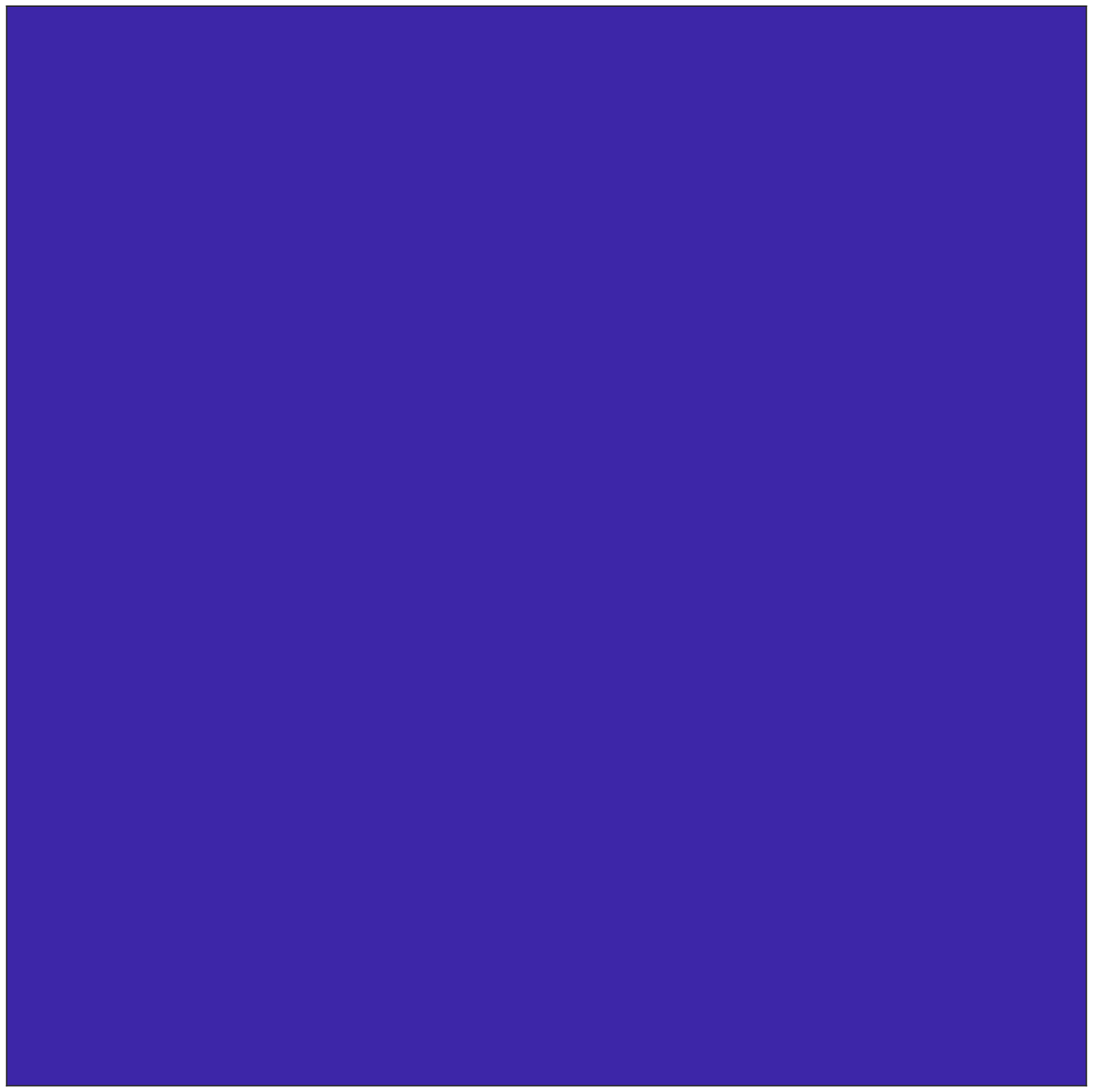}}
\subfigure[GSA \cite{adpativecs}]{
\includegraphics[width=29mm]{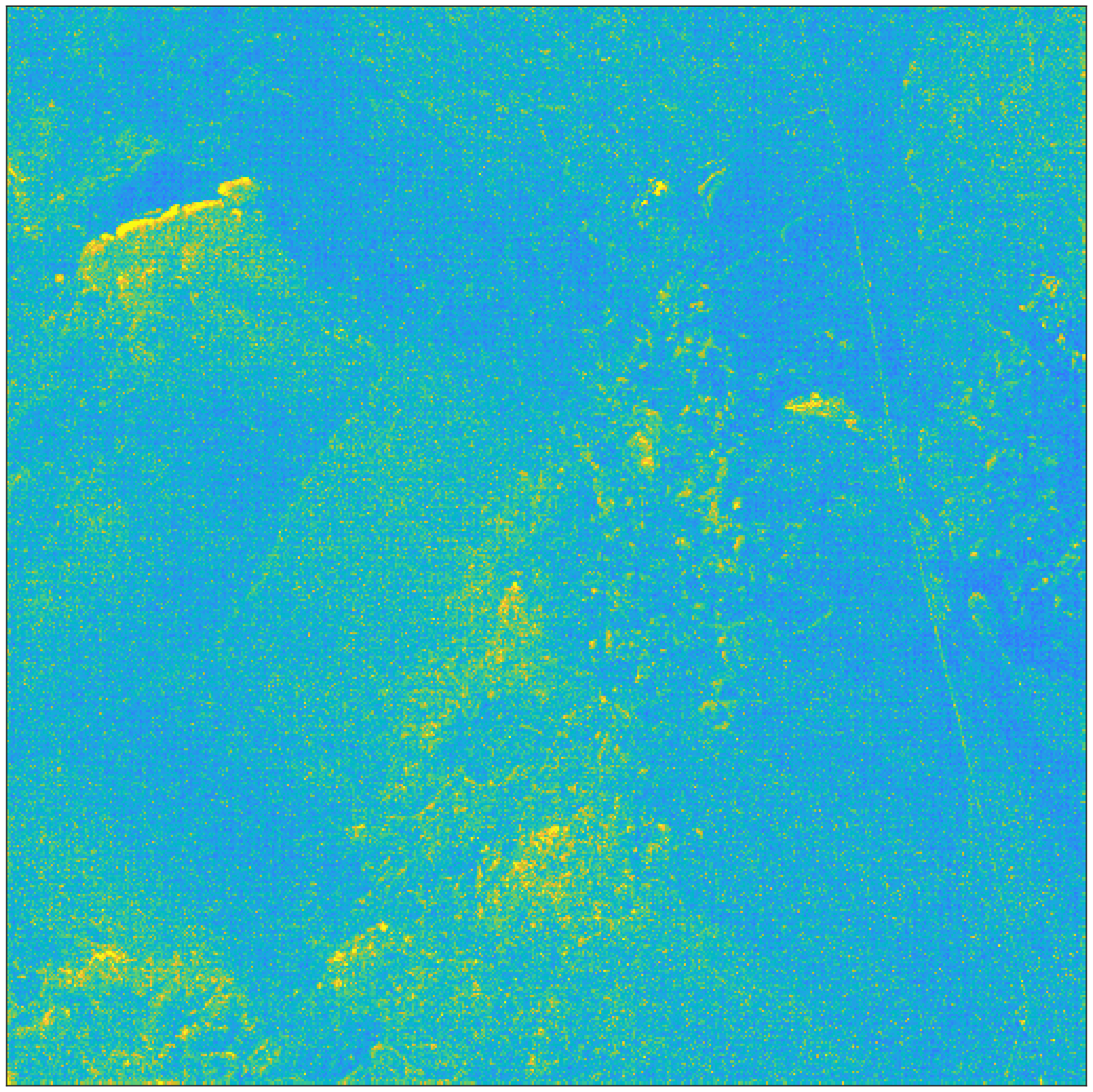}}
\subfigure[GLP-HS\cite{mtfglp}]{
\includegraphics[width=29mm]{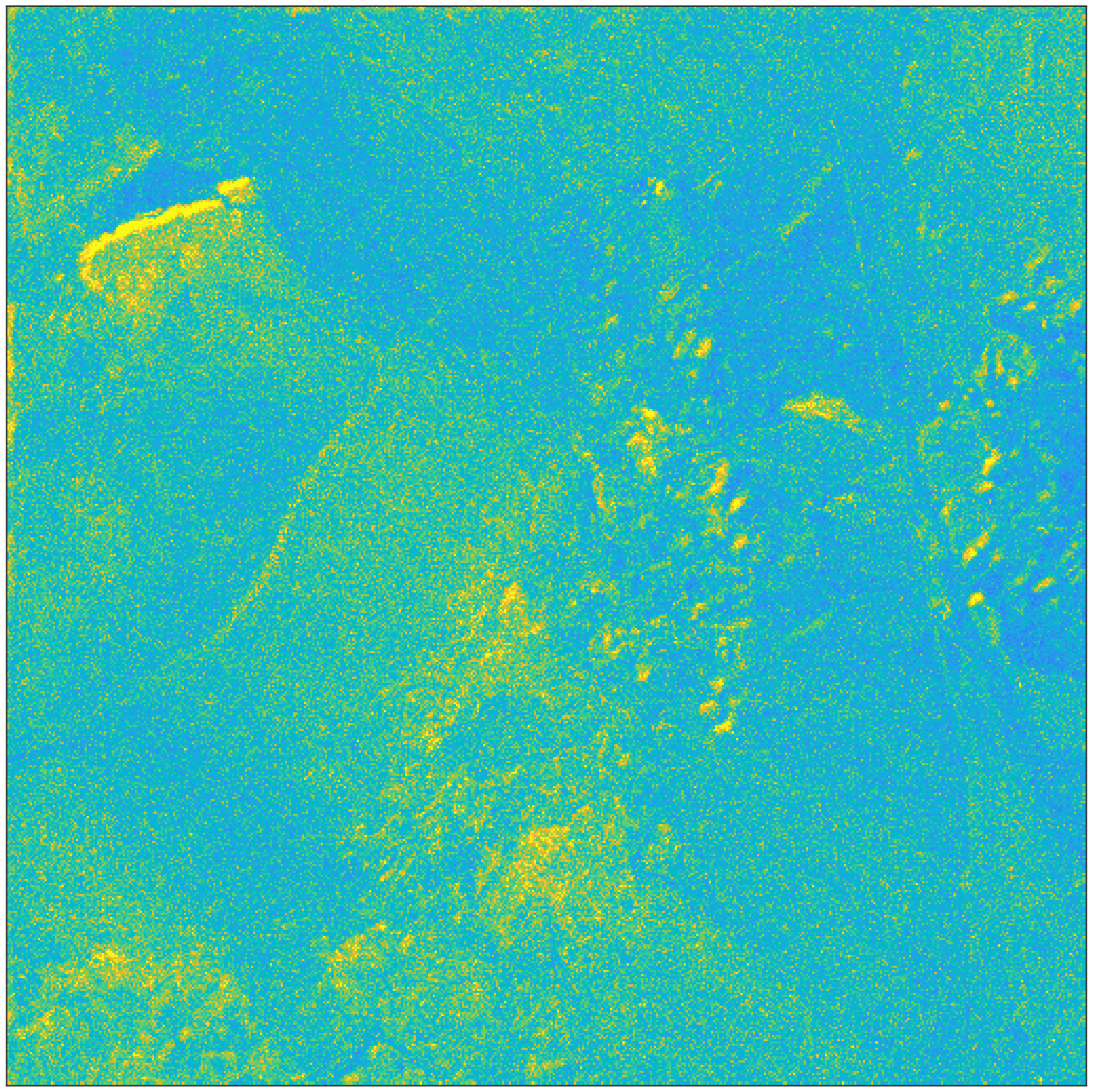}}
\subfigure[NSSR \cite{nssr}]{
\includegraphics[width=29mm]{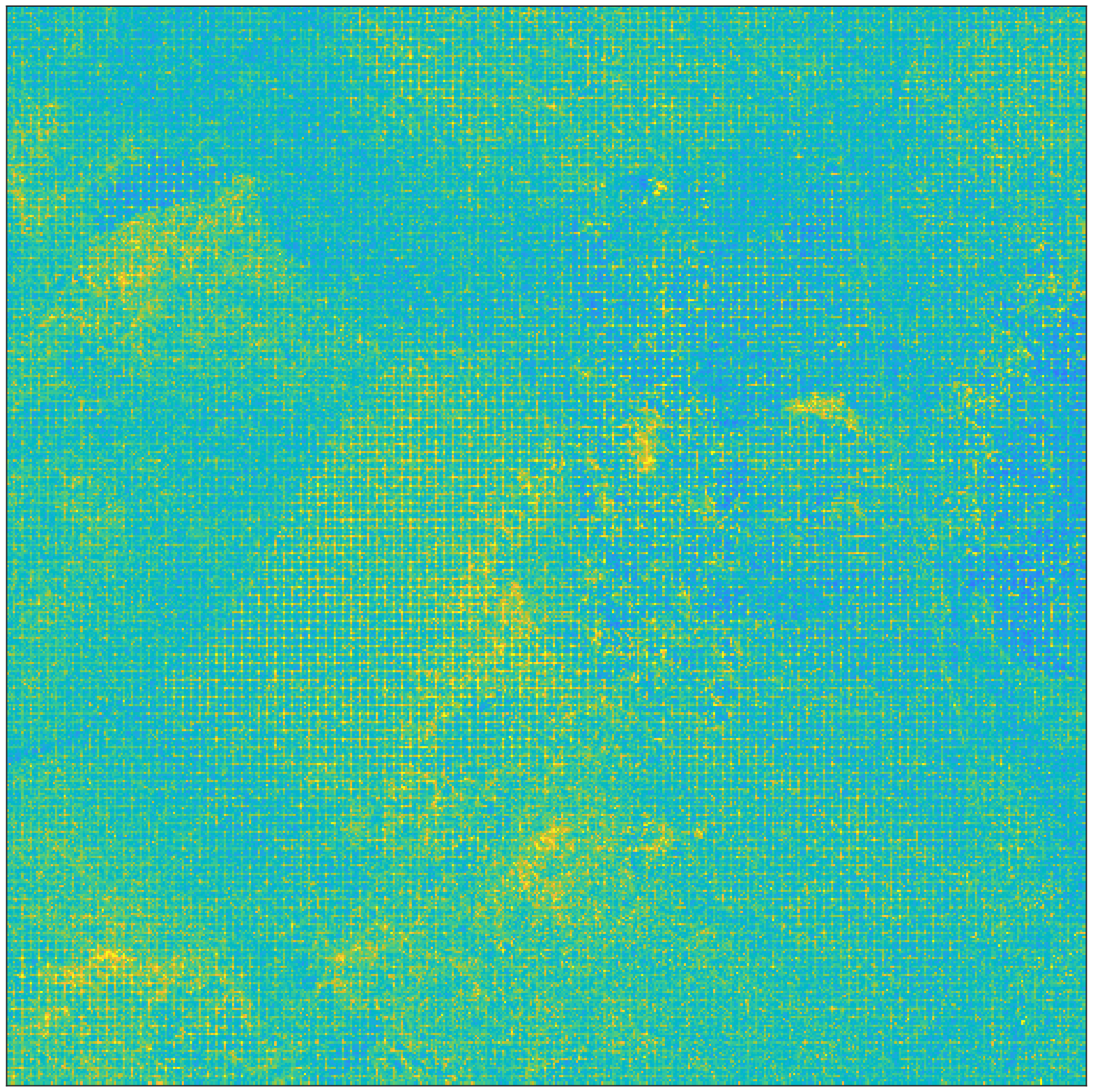}}
\subfigure[CNMF   \cite{cnmf}]{
\includegraphics[width=29mm]{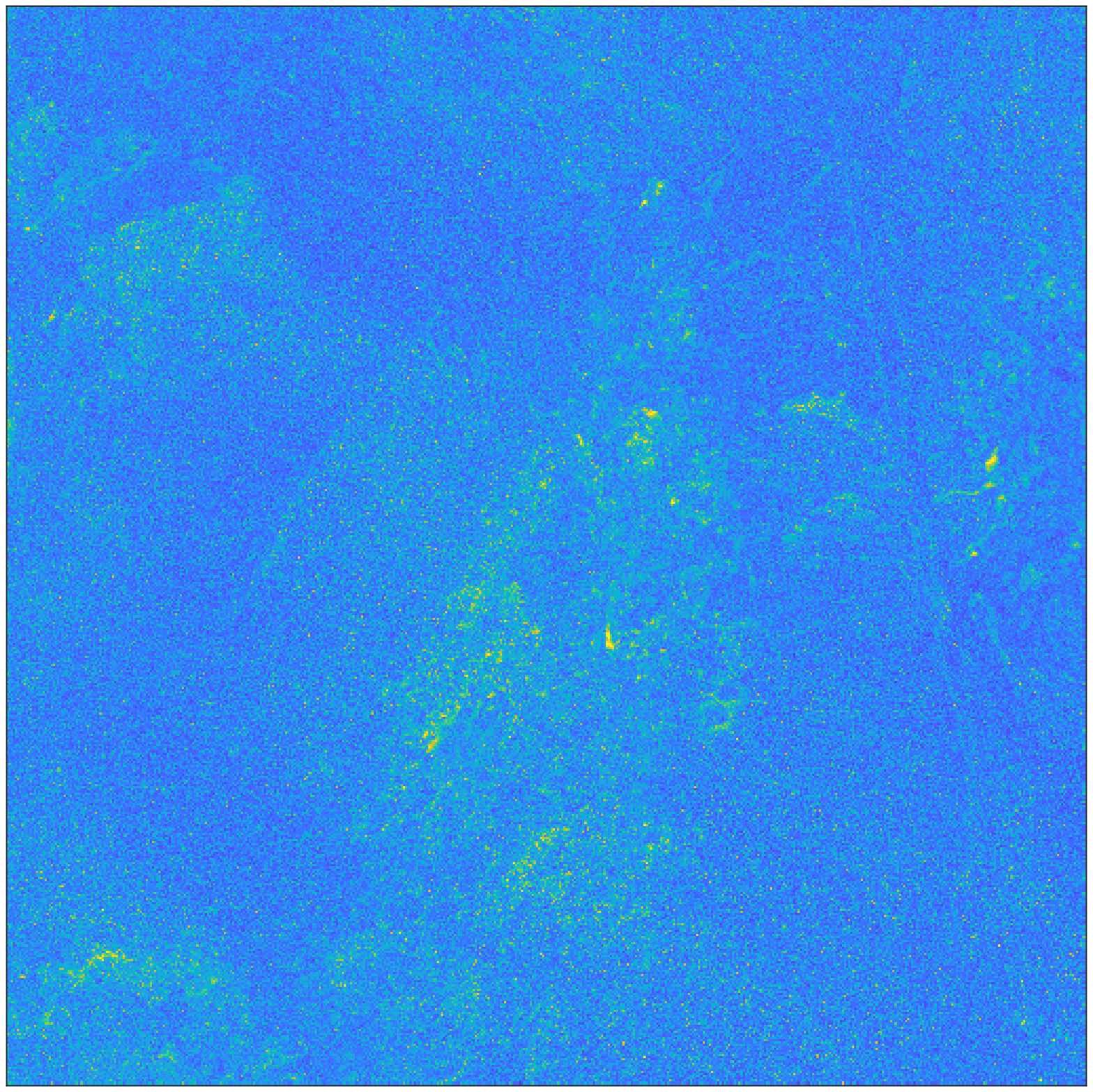}}\\
\subfigure{
\includegraphics[width=29mm]{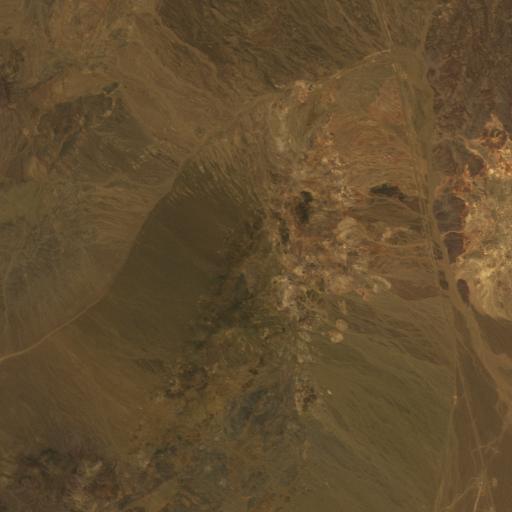}}
\subfigure{
\includegraphics[width=29mm]{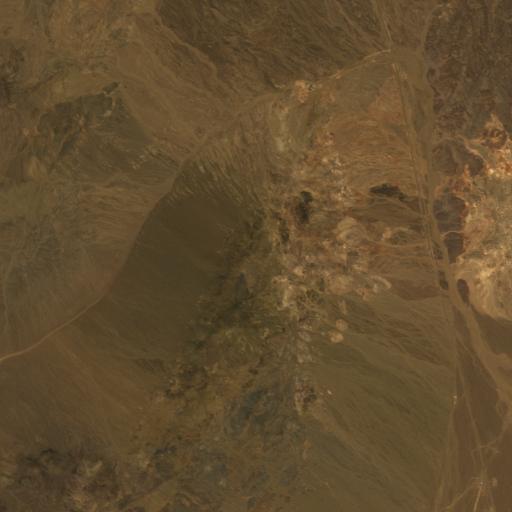}}
\subfigure{
\includegraphics[width=29mm]{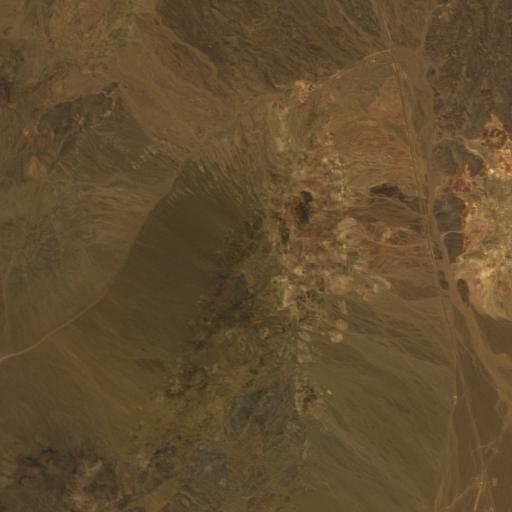}}
\subfigure{
\includegraphics[width=29mm]{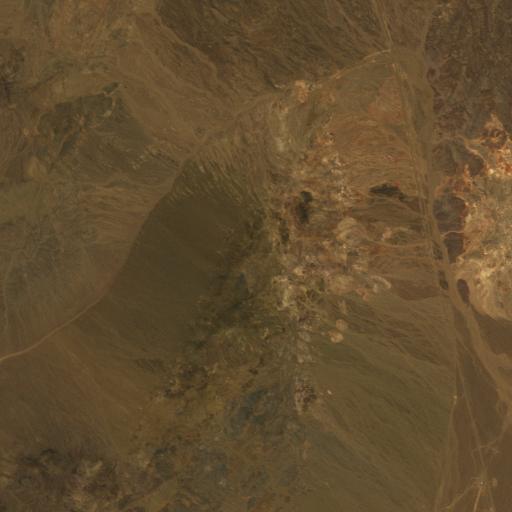}}
\subfigure{
\includegraphics[width=29mm]{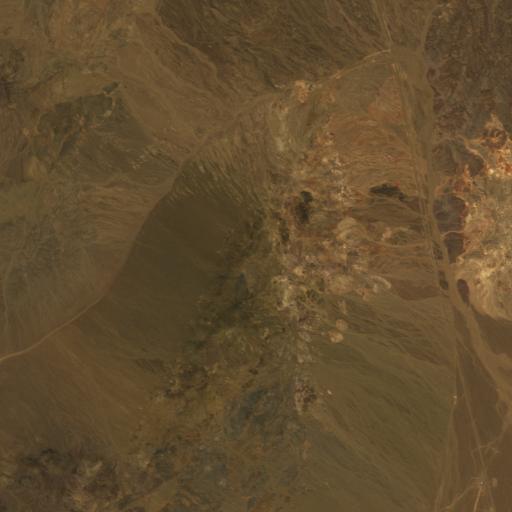}}\\
\setcounter{subfigure}{5}
\subfigure[CSU \cite{csu}]{
\includegraphics[width=29mm]{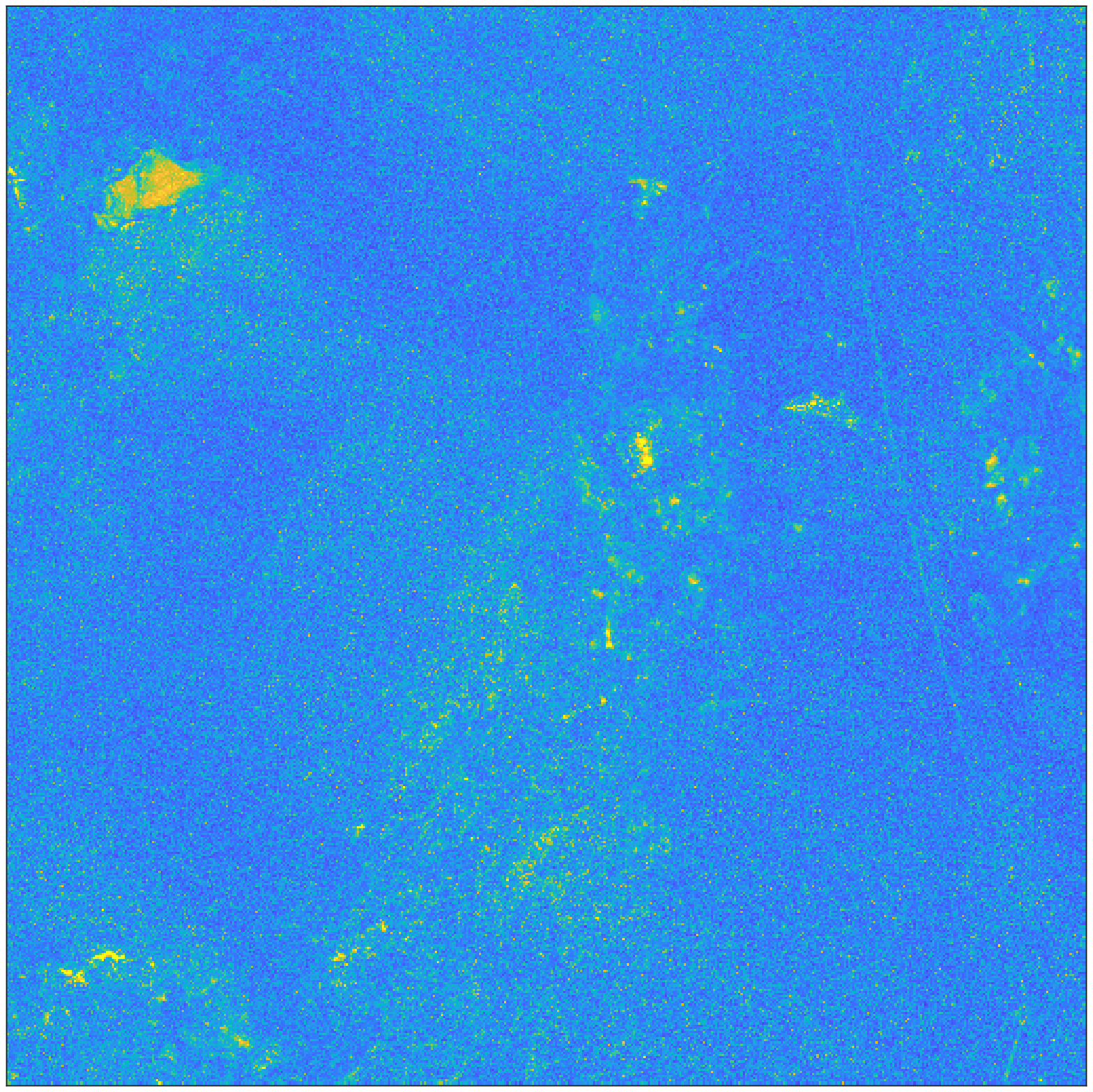}}
\subfigure[Fuse-S   \cite{fuses}]{
\includegraphics[width=29mm]{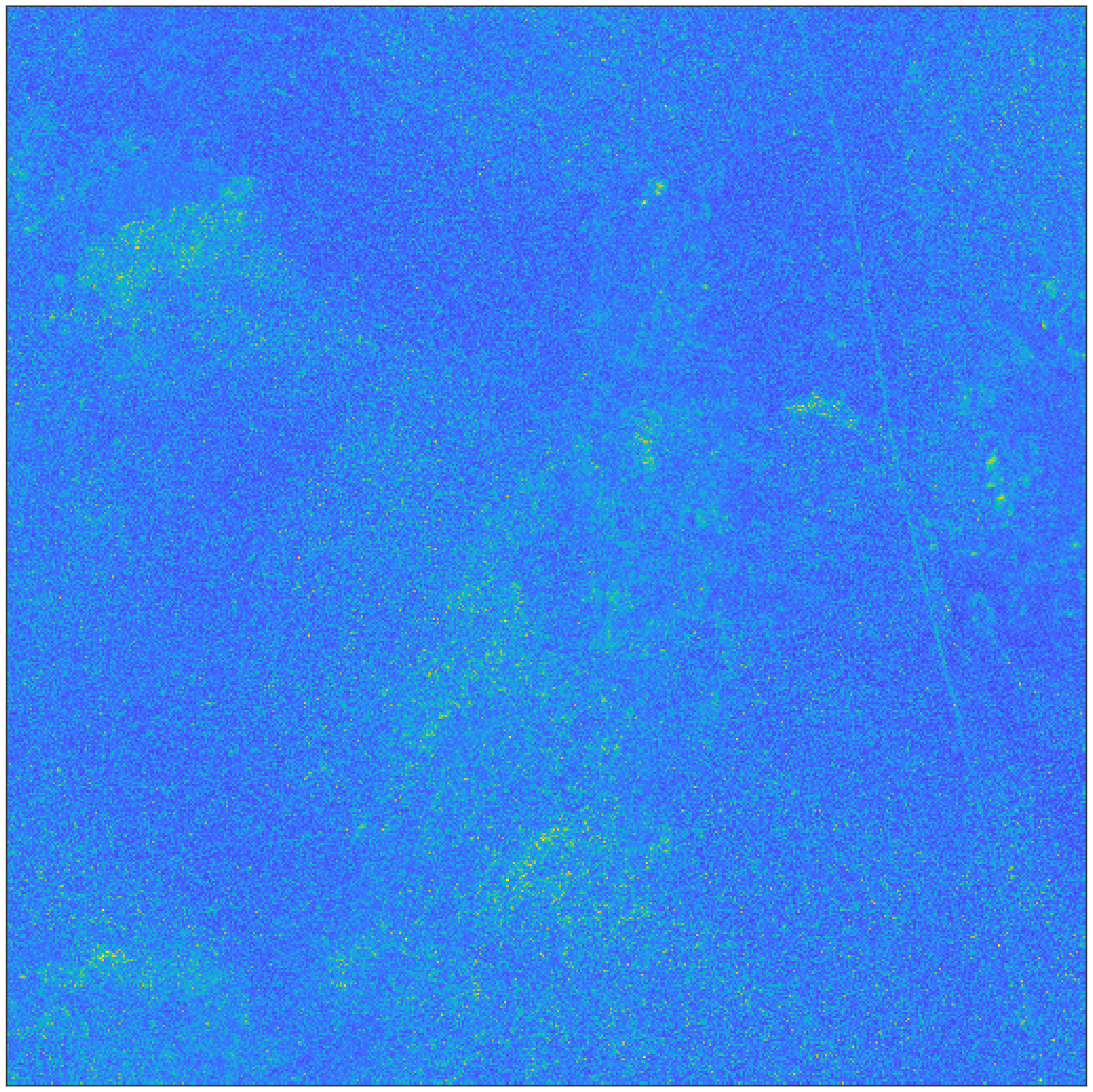}}
\subfigure[CSTF   \cite{cstf}]{
\includegraphics[width=29mm]{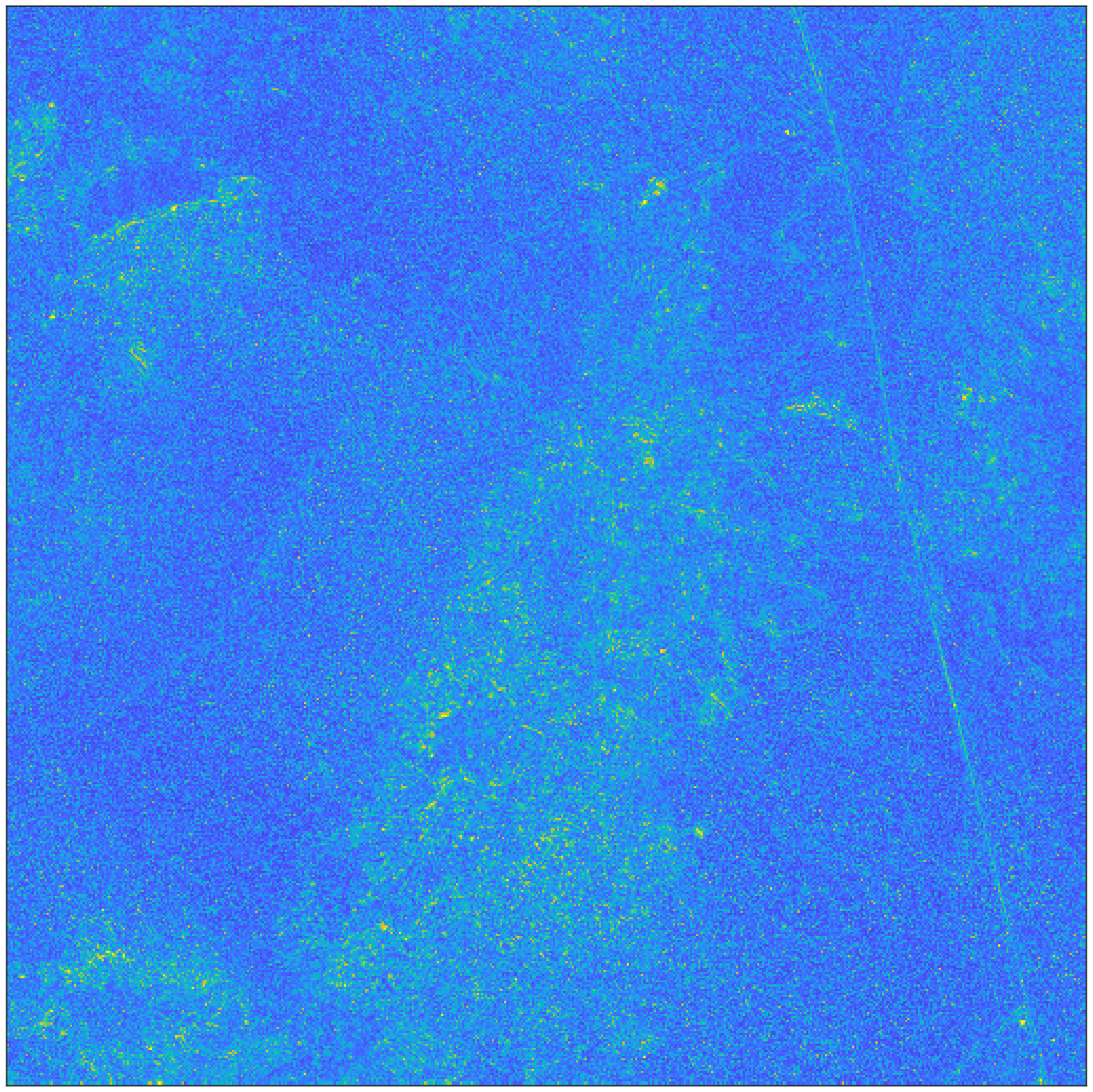}}
\subfigure[LTMR   \cite{ltmr}]{
\includegraphics[width=29mm]{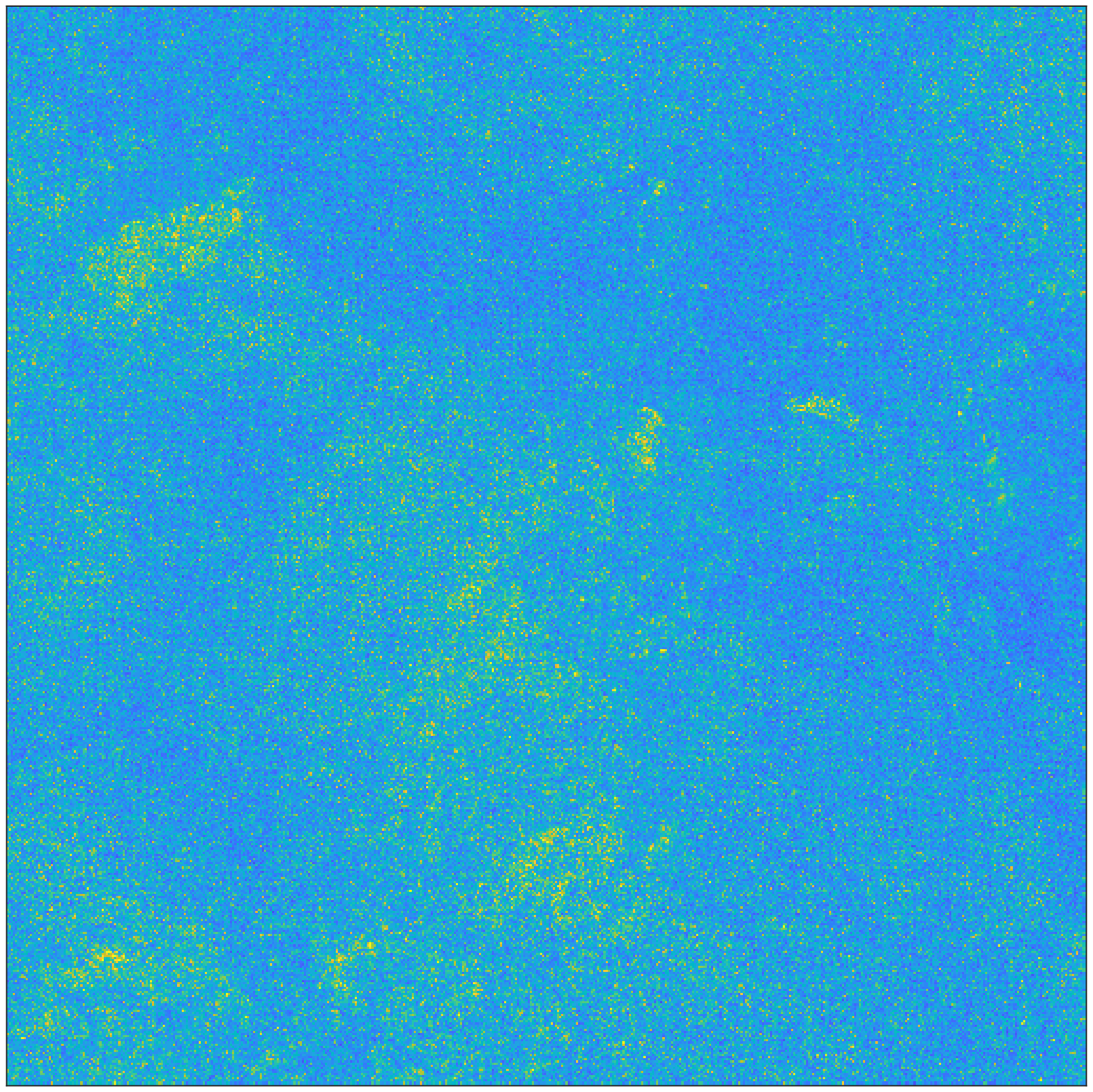}}
\subfigure[CNN-Fus \cite{cnnfus}]{
\includegraphics[width=29mm]{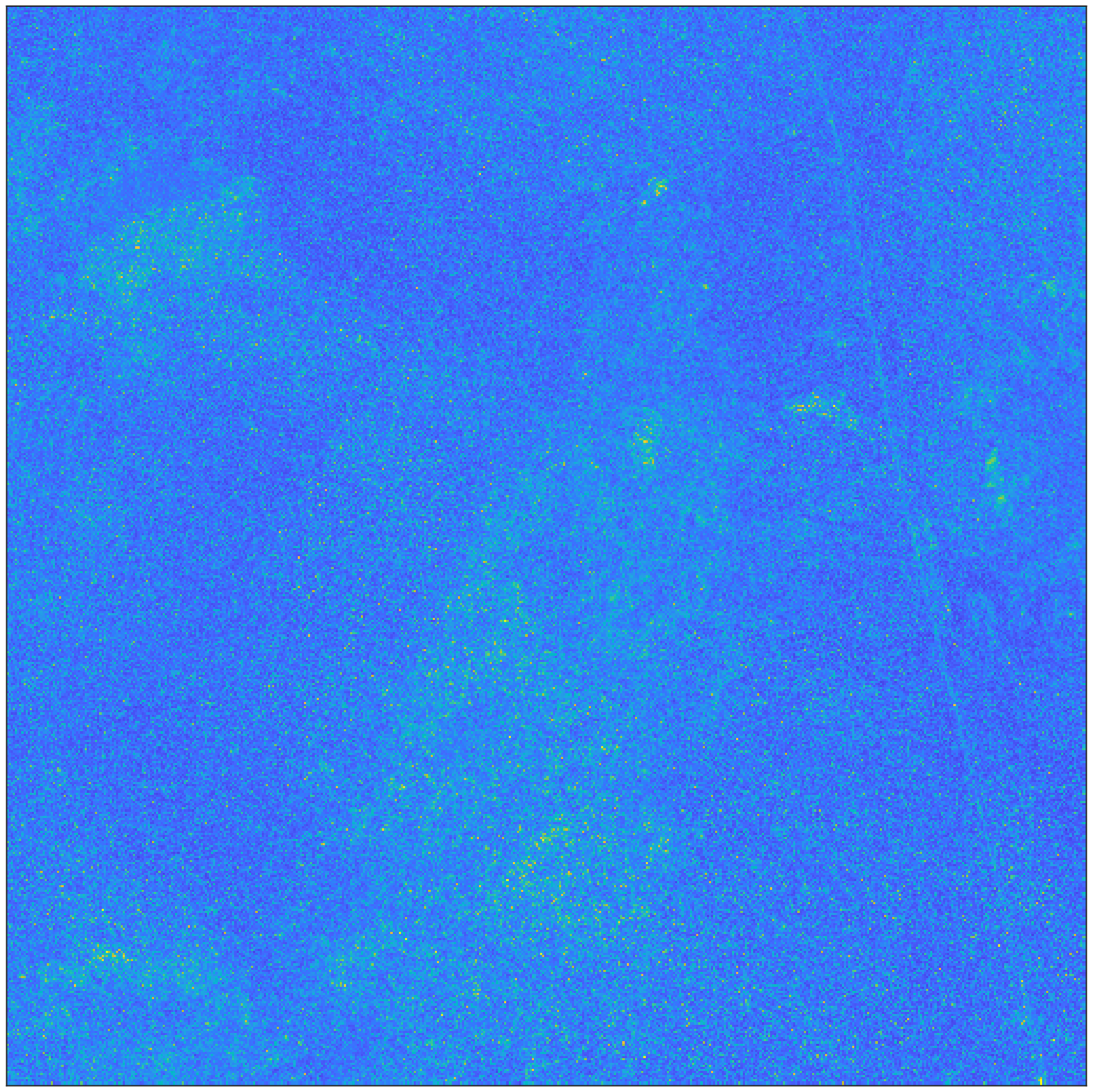}}\\
\subfigure{
\includegraphics[width=120mm]{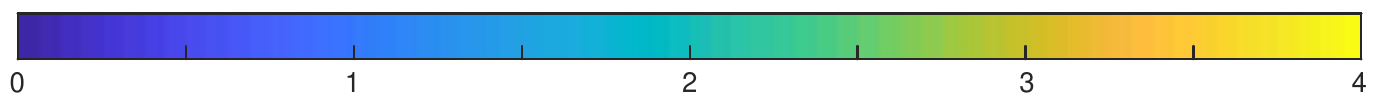}}
\caption{ False color images formed by 30-th,  18-th, and 5-th bands and spectral error images of fused Cuprite Mine by testing methods.
 (a) Reference image. (b) GSA \cite{adpativecs}. (c) GLP-HS\cite{mtfglp}. (d) NSSR \cite{nssr}. (e) CNMF   \cite{cnmf}.  (f) CSU \cite{csu}.   (f) CSU \cite{csu}. (g) Fuse-S   \cite{fuses}. (h) CSTF   \cite{cstf}.  (i) LTMR   \cite{ltmr}.  (j) LTMR   \cite{cnnfus}.   }\label{figcuprite}
\end{figure*}

\begin{figure*}[!t]
\centering
\subfigure{
\includegraphics[width=22mm]{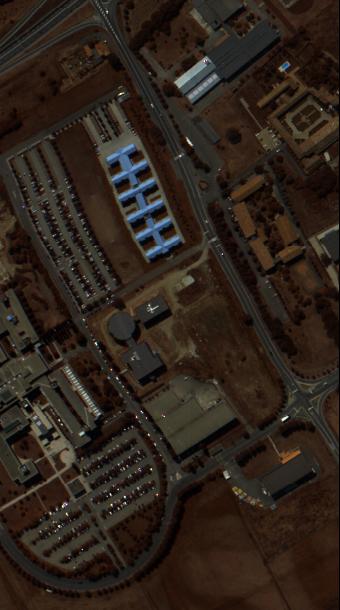}}
\subfigure{
\includegraphics[width=22mm]{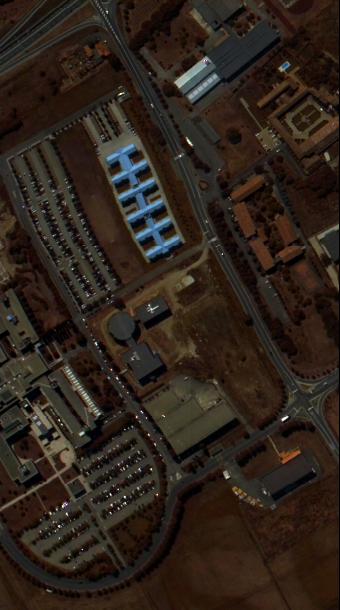}}
\subfigure{
\includegraphics[width=22mm]{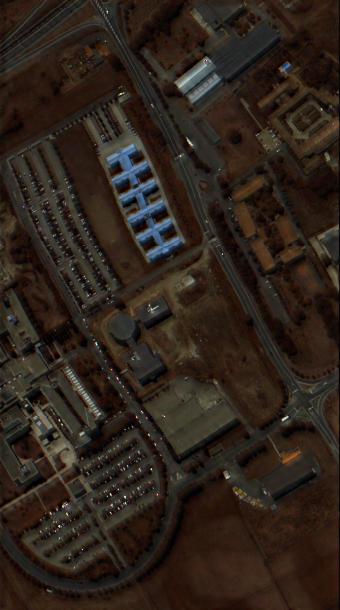}}
\subfigure{
\includegraphics[width=22mm]{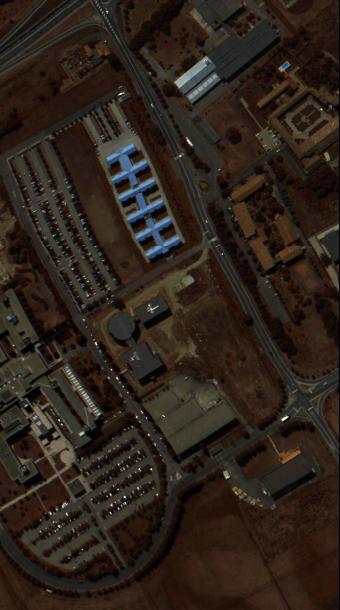}}
\subfigure{
\includegraphics[width=22mm]{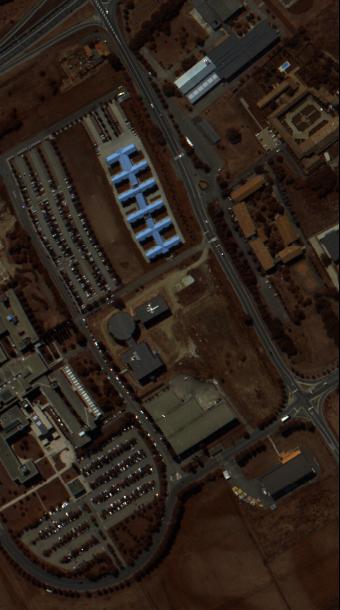}}\\
\setcounter{subfigure}{0}
\subfigure[Reference image]{
\includegraphics[width=22mm]{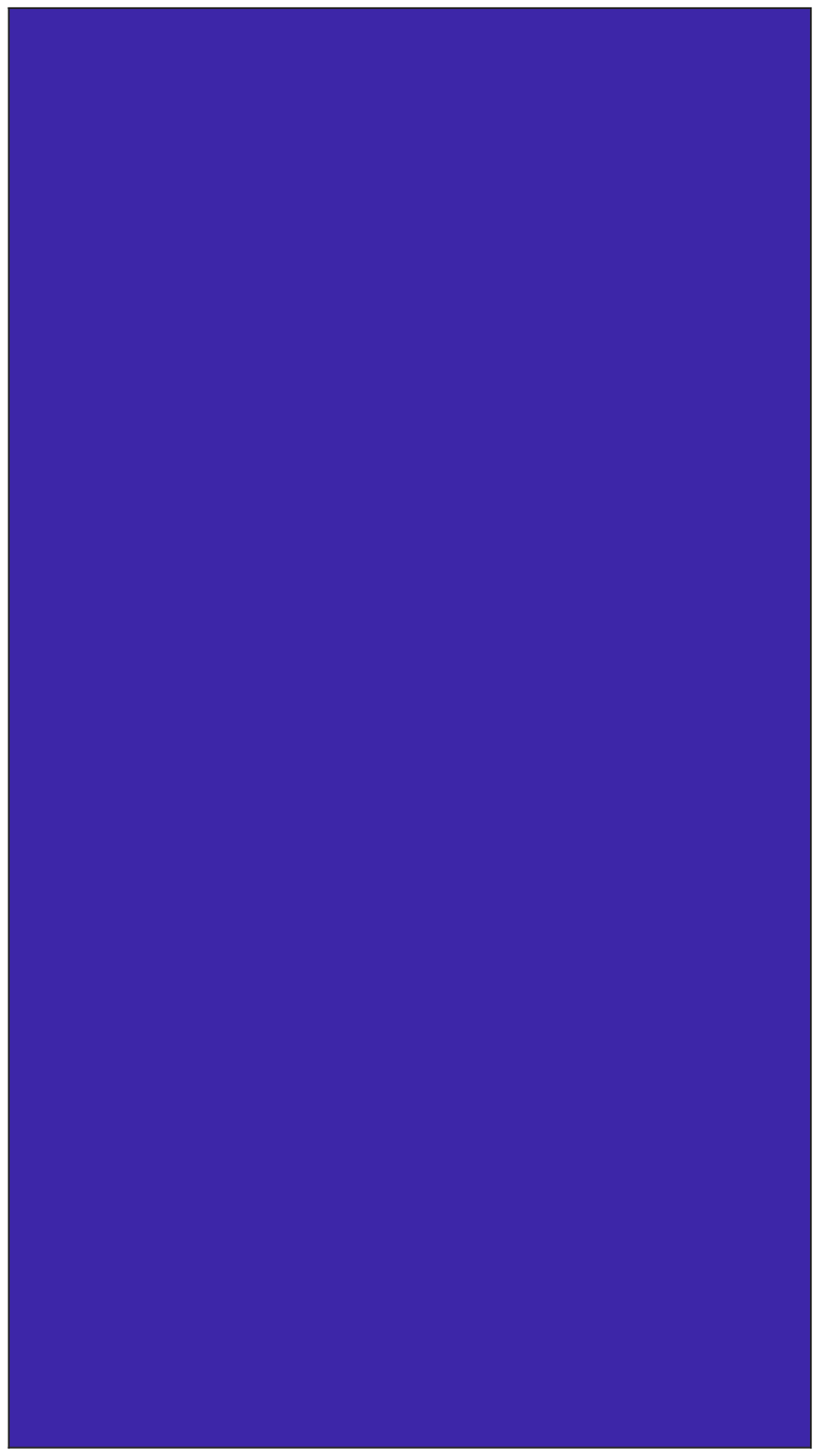}}
\subfigure[GSA \cite{adpativecs}]{
\includegraphics[width=22mm]{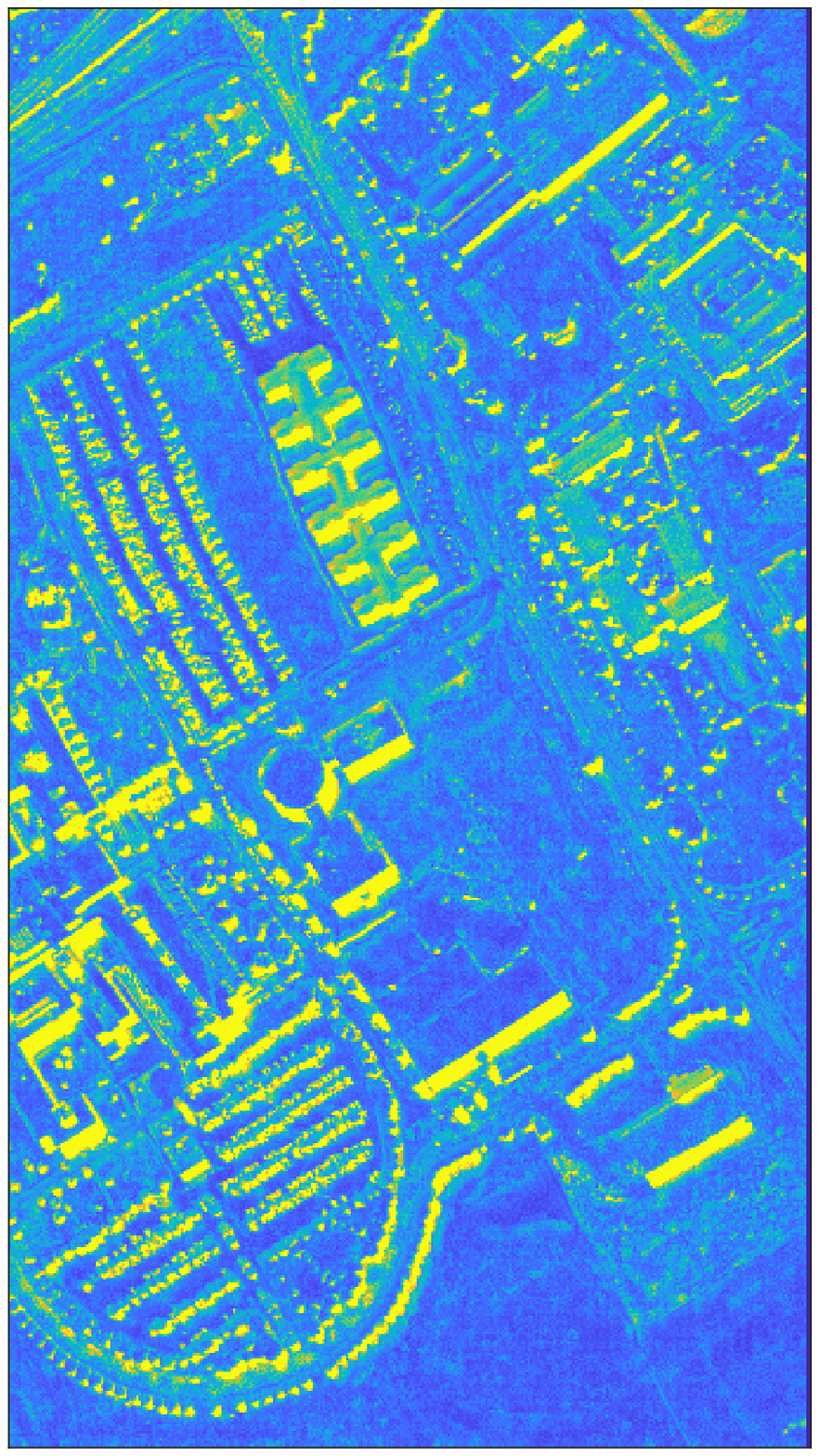}}
\subfigure[GLP-HS\cite{mtfglp}]{
\includegraphics[width=22mm]{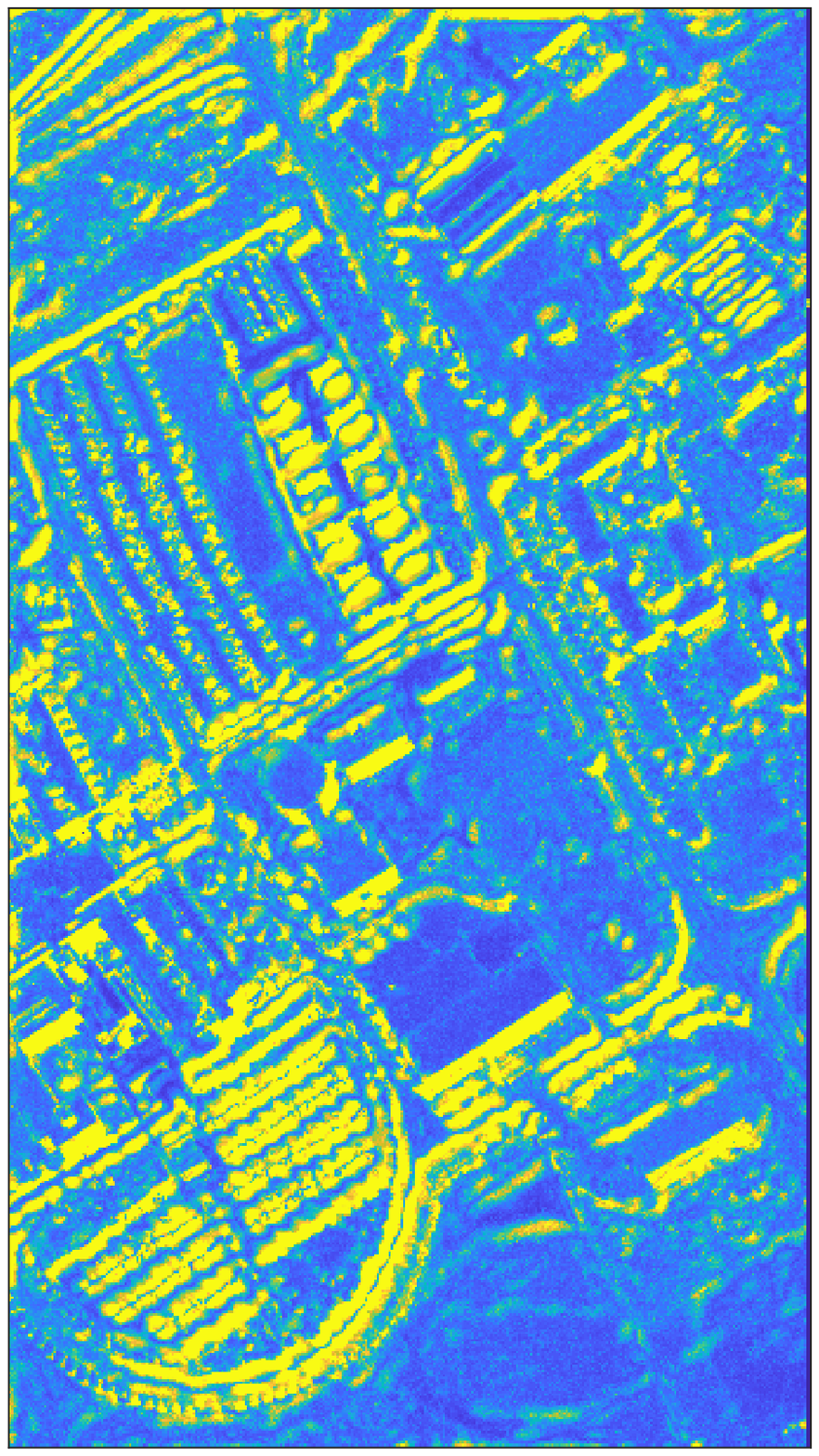}}
\subfigure[NSSR \cite{nssr}]{
\includegraphics[width=22mm]{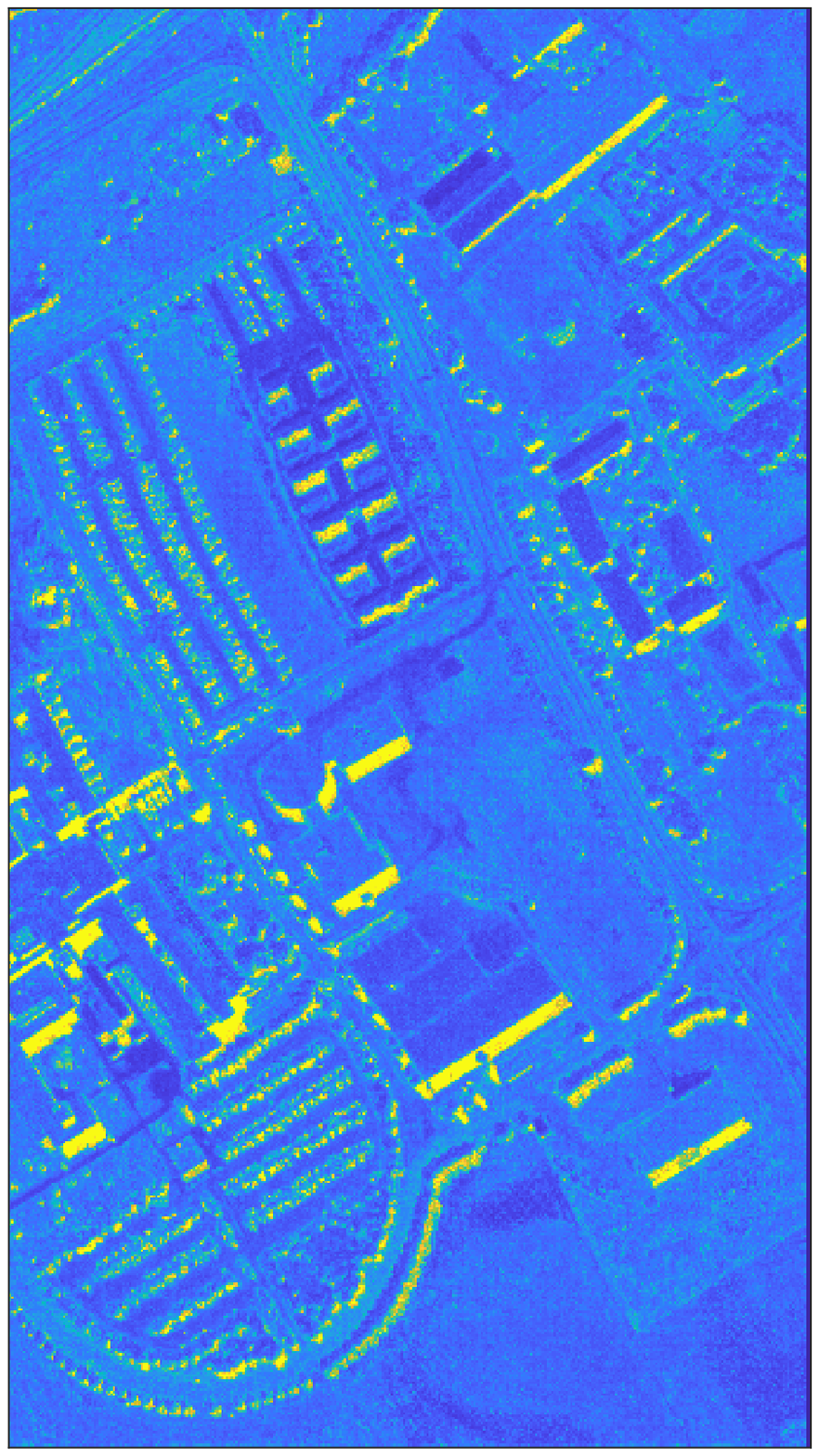}}
\subfigure[CNMF   \cite{cnmf}]{
\includegraphics[width=22mm]{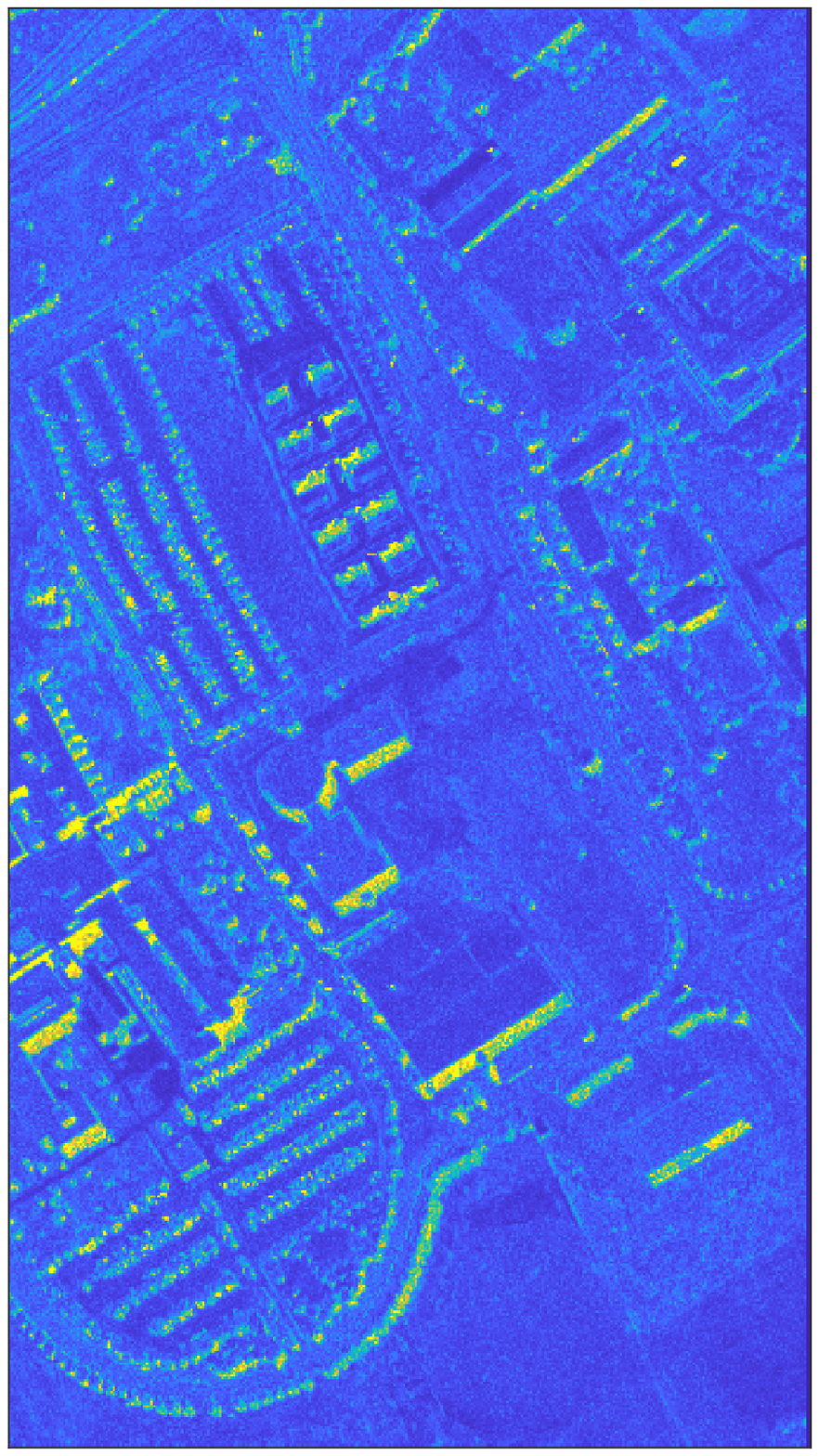}}\\

\subfigure{
\includegraphics[width=22mm]{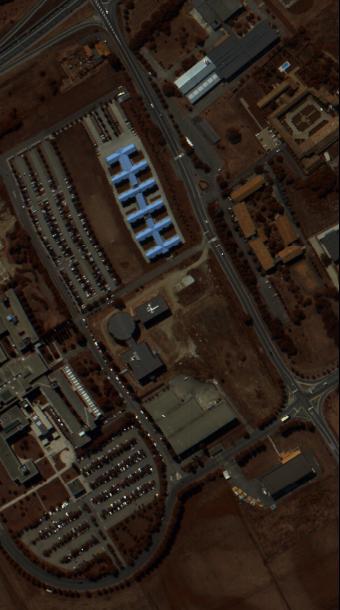}}
\subfigure{
\includegraphics[width=22mm]{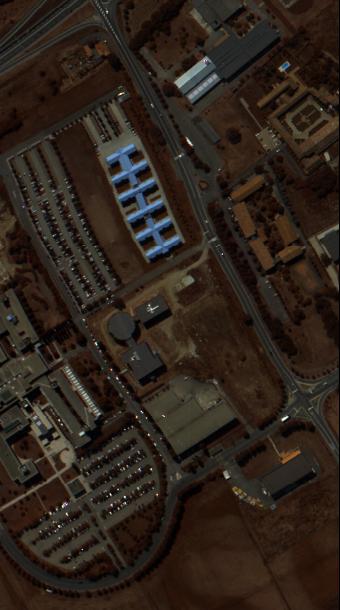}}
\subfigure{
\includegraphics[width=22mm]{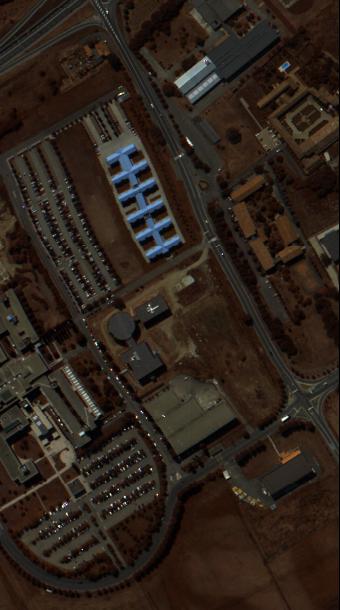}}
\subfigure{
\includegraphics[width=22mm]{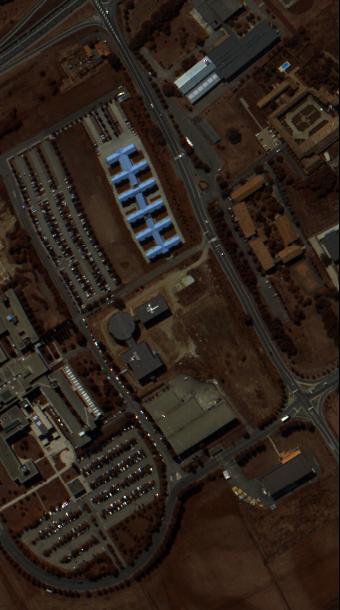}}
\subfigure{
\includegraphics[width=22mm]{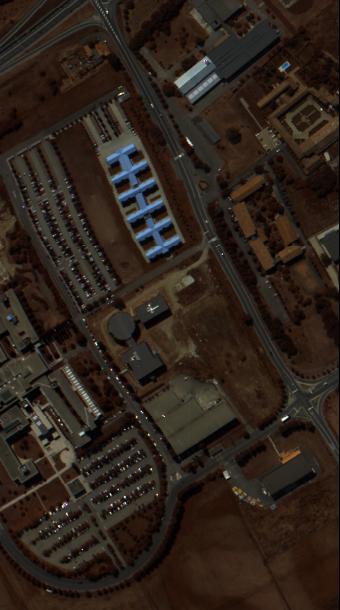}}\\
\setcounter{subfigure}{5}
\subfigure[CSU \cite{csu}]{
\includegraphics[width=22mm]{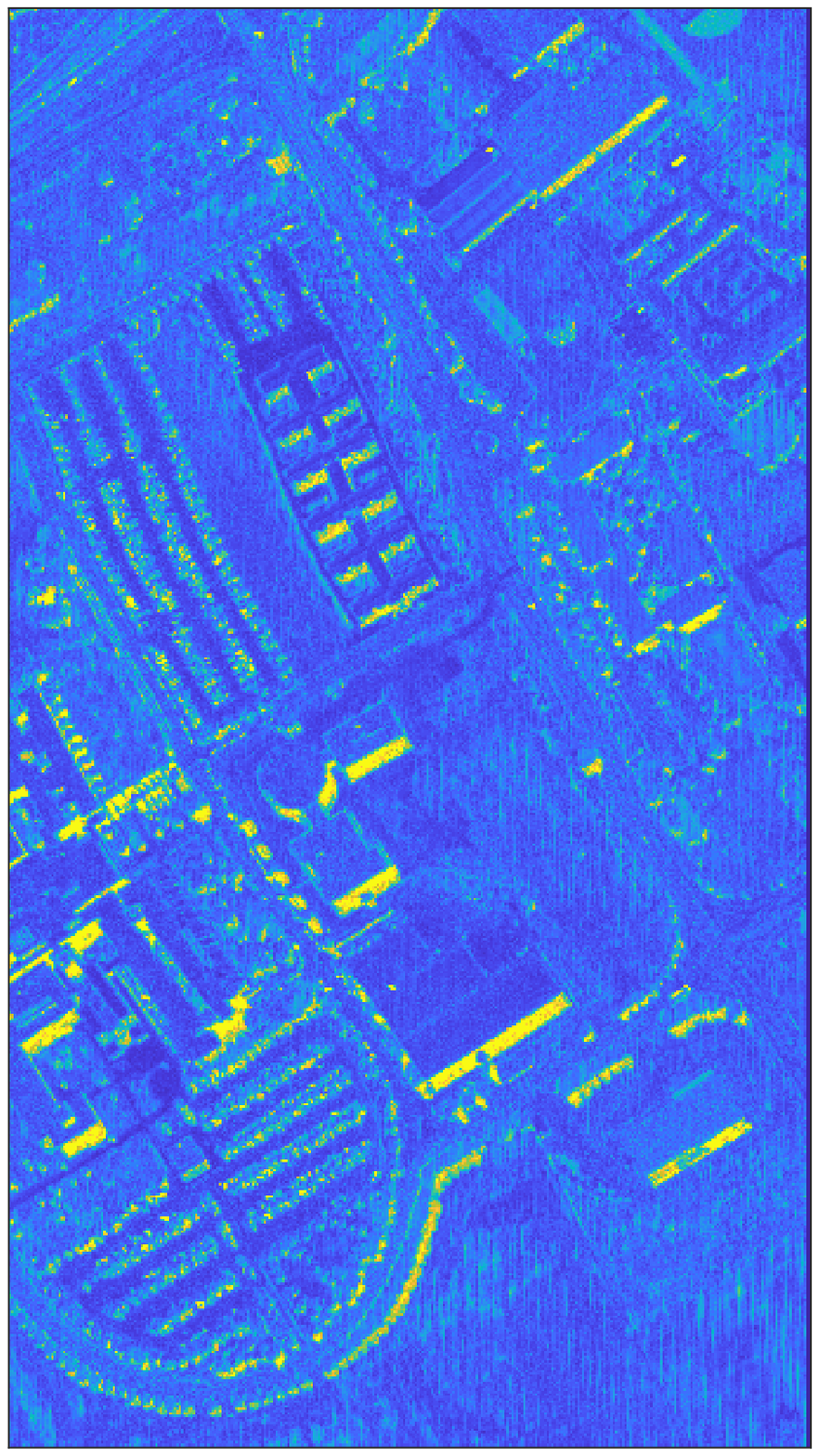}}
\subfigure[Fuse-S   \cite{fuses}]{
\includegraphics[width=22mm]{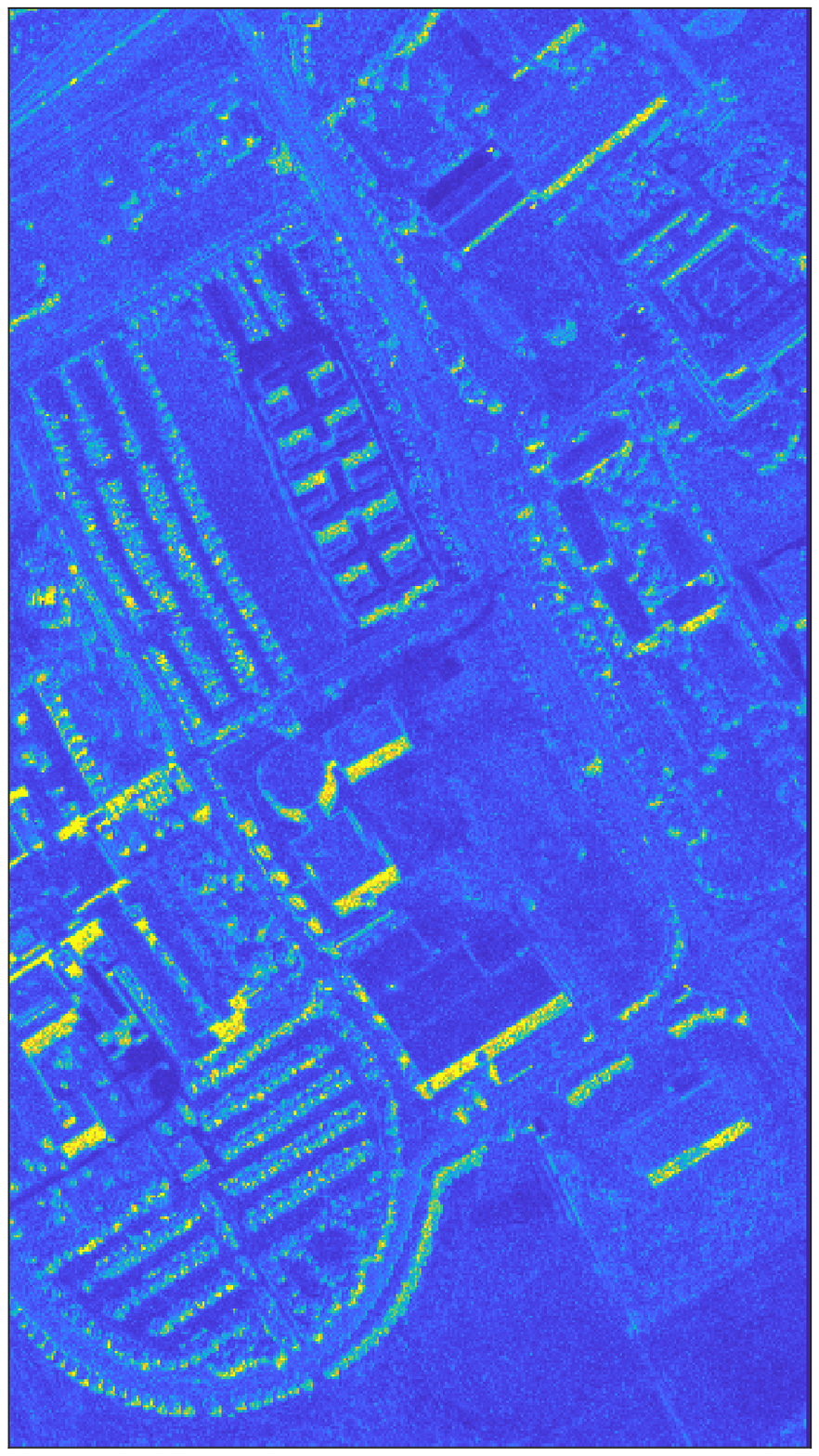}}
\subfigure[CSTF   \cite{cstf}]{
\includegraphics[width=22mm]{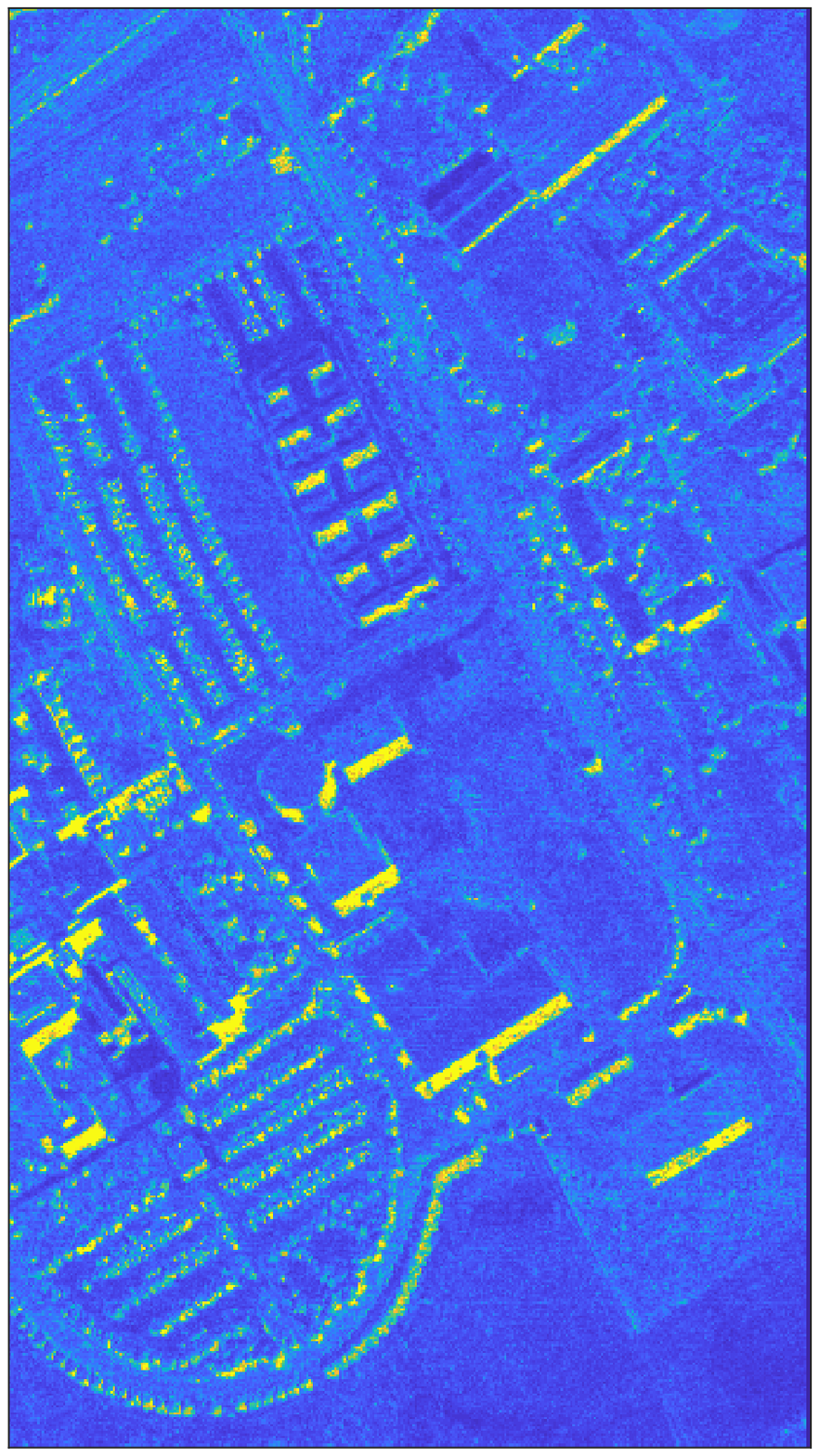}}
\subfigure[LTMR   \cite{ltmr}]{
\includegraphics[width=22mm]{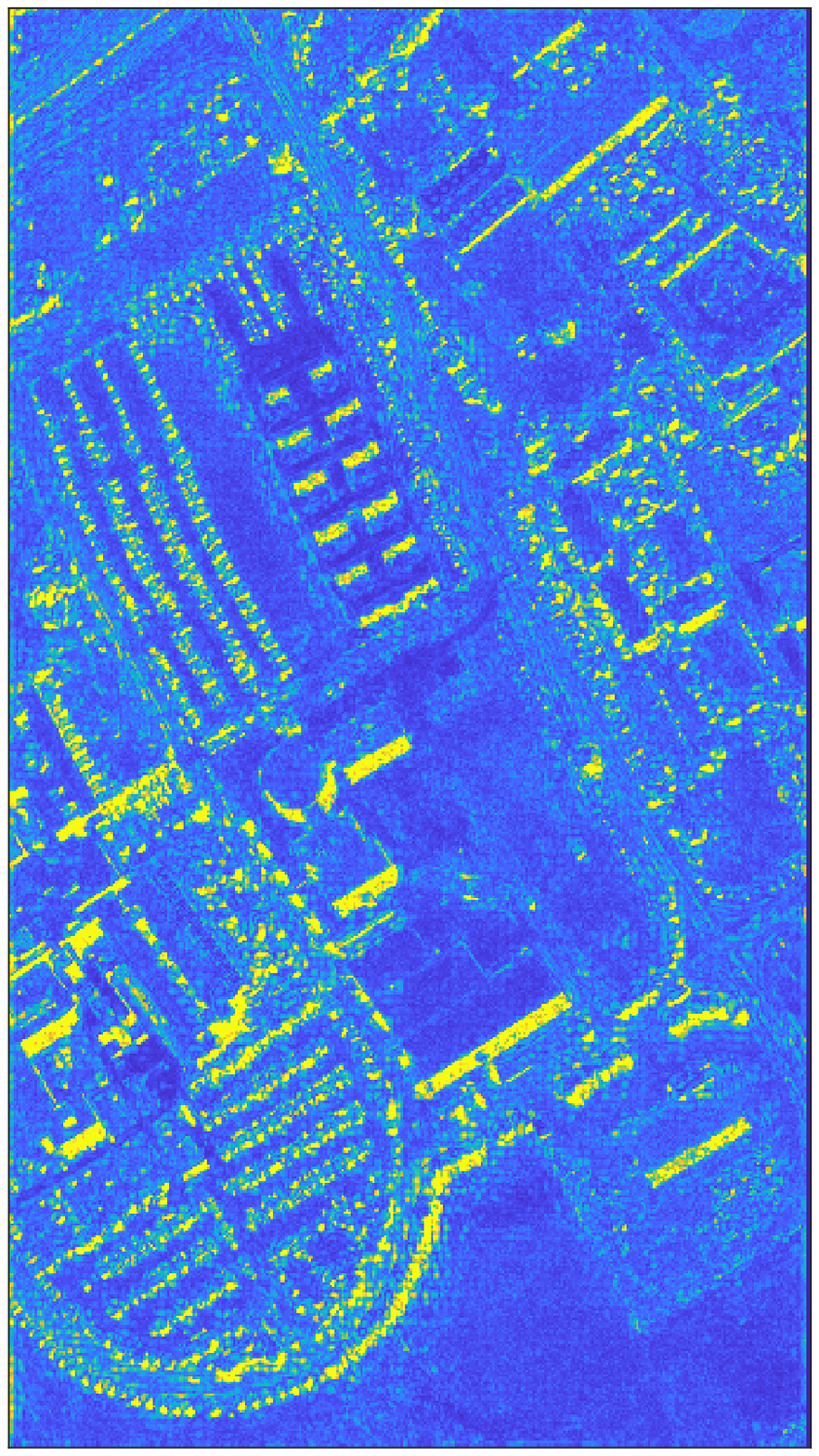}}
\subfigure[CNN-Fus \cite{cnnfus}]{
\includegraphics[width=22mm]{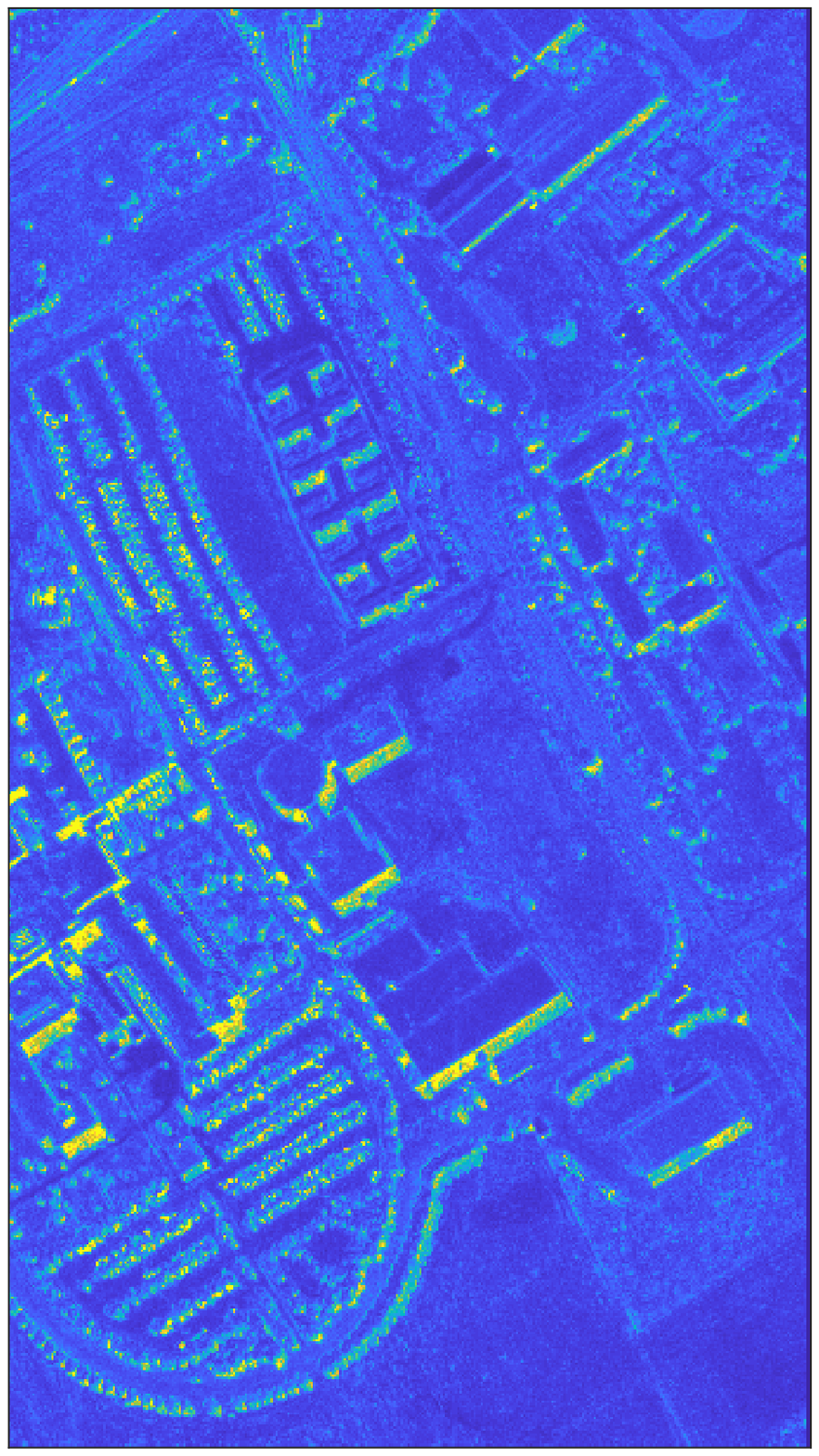}}\\
\subfigure{
\includegraphics[width=110mm]{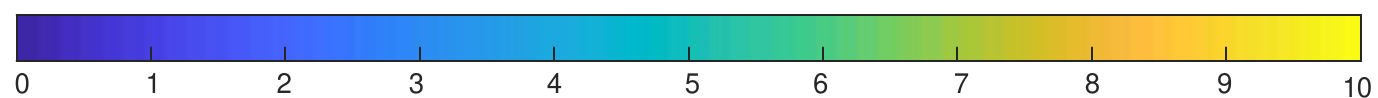}}
\caption{ False color images formed by 60-th,  29-th, and  7-th bands and spectral error images of fused Pavia University by testing methods.
 (a) Reference image. (b) GSA \cite{adpativecs}. (c) GLP-HS\cite{mtfglp}. (d) NSSR \cite{nssr}. (e) CNMF   \cite{cnmf}.  (f) CSU \cite{csu}.   (f) CSU \cite{csu}. (g) Fuse-S   \cite{fuses}. (h) CSTF   \cite{cstf}.  (i) LTMR   \cite{ltmr}.  (j) LTMR   \cite{cnnfus}.   }\label{figpavia}
\end{figure*}

\subsection{Experimental Dataset}

\subsubsection{Pavia University}
Pavia university is obtained by ROSIS-3 optical airborne sensor over the area  of the University of Pavia, Italy. The HSI consists of $610\times340$ pixels and has a spatial resolution of 1.3m. It has 115 bands, and 93 spectral bands are preserved after removing the spectral bands of low signal-noise-ratio (SNR).  To simulate the low-resolution HSI, we apply an $7\times7$ Gaussian blur kernel with a mean value of zero and a standard deviation of 3 to filter the reference image and then subsample the filtered image with factor 5. The four-band MSI is simulated by filtering  the reference image with  the IKONOS-like  spectral response. The independent and identically distributed Gaussian noise  is added to the simulated MSI (35dB) and simulated HSI (32dB).

\subsubsection{Cuprite Mine}
 Cuprite Mine is obtained by the AVIRIS sensor  over
 Cuprite mining district in Nevada in 1995. This image contains $512\times614$ pixels, and top-left $512\times512$ pixels are used for the experiments. The spatial resolution of this image is 17m.
 This image has 224 spectral bands, 188 spectral bands are preserved after removing the spectral bands of low signal-noise-ratio (SNR). To simulate the low-resolution HSI, we apply  an $7\times7$ Gaussian blur kernel with a mean value of zero and a standard deviation of 3 to filter the reference image and then subsample the filtered image with factor 4.  Six-band high-resolution MSI is simulated by selecting wavelengths 480, 560, 660, 830, 1650, and 2220nm.  The  independent and identically distributed Gaussian noise  is added to the simulated MSI (35dB) and simulated HSI (32dB).

\subsection{Quantitative Evaluation}
We further evaluate the quality fused HSI obtained by the testing approaches via quantitative indexes, consisting of PSNR, ERGAS, SAM, UIQI, and T. Tables \ref{tabcuprite} and  \ref{tabpavia}  show the quantitative indexes on  Cuprite Mine and  Pavia University, respectively.
From  two tables, the following observations can be obtained:\\
1) The pan-sharpening based HSI-MSI fusion methods, including GSA and GLP-HS, have a low computational burden and can be implemented much faster than other methods. However, the quality metrics of them are  comparatively poor. The underlying reason is that they do not consider the observational model of the MSI and HSI, which is crucial for the fusion.\\
2) The MF based approaches, consisting of NSSR, CNMF, CSU, and Fuse-S, achieve good performance on the two datasets. Among these methods, Fuse-S achieves the best performance, which exploits spectral subspace representation and patch-based sparse prior. However, it is very time-consuming, the computational burden mainly comes from the patch-based dictionary learning and sparse coding. The CNMF can achieve
good performance and high computational efficiency among the MF based approaches.
\\
3) In the TR based fusion methods, the CSTF and LTMR produce better quality indexes on Cuprite and Pavia University. It is because that the LTMR is based on spatial similarities of the fused HSI. The spatial resolution of Cuprite Mine is  lower than that of Pavia University, and therefore that  non-local spatial similarities are also weaker. In general, the TR based approaches are also comparatively  time-consuming.\\
4)  CNN-Fus is deep CNN based fusion approaches and achieves the best fusion results on most of the quality metrics for the two datasets. The advantages of this method mainly come from spectral subspace representation and CNN  based  coefficients estimation, in which the spectral subspace representation can well maintain the spectral characteristics of the HSI, and CNN  is good at preserving the spatial structures of the image.

\subsection{Visual Evaluation}
In addition to the quality metric evaluation, we also visually evaluate the fused images. Figure \ref{figcuprite} shows   false-color images formed by 30-th,  18-th, and 5-th bands and spectral error images of fused Cuprite Mine by testing methods, in which the spectral error image reflects the SAM for the fused HSI, and the false-color images mainly reflect the spatial structures of the fused HSI. As can be seen from Figure \ref{figcuprite},
the  fused HSIs obtained by GSA, GLP-HS, and NSSR have severe spectral distortions, and the HSIs fused by CNMF,
Fuse-S, CSTF, and CNN-Fus have much less spectral distortions.
 Figure  \ref{figpavia} shows   false-color images formed by 60-th,  29-th, and 7-th bands and spectral error images of fused Pavia University by testing methods. As can be seen from Figure \ref{figpavia}, the HSIs fused by  GSA and GLP-HS
  contain strong spectral errors. The fused HSIs obtained by  Fuse-s and CNN-Fus have
 less spectral distortions. The advantages of the two methods mainly come from the
 spectral subspace representation, which can effectively model the spectral redundancies
 and similarities.  The HSI fused by CSTF, CNMF, and CSU also contain relatively less
 spectral distortions, since they also make use of the low-rank prosperities of the spectral mode.

\section{Challenges and New Guidelines for HSI-MSI fusion}
   HSI-MSI fusion has made significant progress in the past ten years. However, it remains some challenges. We present the existing challenges and new guidances for HSI-MSI fusion.

\subsection{Image Registration}
   The image registration intends to
 geometrically align two images of the same scenario acquired  by different sensors or at different times.
   Since HSI-MSI fusion requires that the MSI and HSI capture the image of
   the same scenario, and therefore  image registration   is a very crucial
    pre-processing step for HSI-MSI fusion.
      The image registration methods are categorized as area-based \cite{area} and feature-based methods \cite{sift}. Although image registration approaches, such as  scale-invariant
Fourier transform \cite{sift}, can produce   most of the true matches, and it remains some false matches. However, most of the fusion methods assume that the HSI and MSI are perfectly aligned, and do not account for the distortions caused by the non-rigid registration. The fusion methods, which consider the non-rigid registration, will be a very important topic in future research. Since the deep CNN based methods solve the fusion problem in a supervised way, they may have superiority to reduce the distortions caused by non-rigid alignment.

\subsection{ HSI-MSI Fusion for Multi-temporal Images}
The remote sensing HSI and MSI of the same scenario are often acquired at different time, which may result in different ground objects in the HSI and MSI. In this case, the HSI-MSI fusion is a very challenging problem, since the observation  models for the HSI and MSI are hard to establish, and the image alignment is also tough.
The existing fusion methods scarcely consider this challenging and meaningful problem, which needs to receive  more attention. When applying the existing methods for the fusion of multi-temporal images, the fused image has obvious flaws and blur in the areas where the  ground objects are changed.
 One possible way to solve the problem is to find the changed areas and giving less weight for the changed areas in the MSI.

\subsection{ HSI-MSI Fusion for Big   Spatial Resolution Differences}
In the Pan-sharpening, the spatial downsampling factor between PAN image and MSI  factor is often 4. However, the spatial downsampling factor for HSI and MSI is often much higher than 4. For example, GF-2 can acquire MSI of 4m GSD, and GF-5 acquire MSI of 30m GSD. When the spatial resolutions of HSI and MSI have big differences, the essence of HSI-MSI fusion is solving a severely ill-posed problem, since most of the spatial information is lost. In this way, the fused HSI may contain severe spatial distortions. Therefore, the HSI-MSI fusion for big   spatial resolution differences is a very challenging problem, and more efforts need to be made to solve this problem. The key to solving this problem is estimating the spatial degradation model accurately, which can  reduce the spatial distortions.

\subsection{The estimation of PSF and SRF}
Many fusion methods, such as MF based methods and TR based ones, highly rely on the observation model \eqref{eq43}, which assumes that the PSF and the SRF of the imaging sensor  known. In the simulated data fusion, these methods can obtain satisfactory fusion results based on the perfectly-known PSF and SRF. However, they may not produce good fusion results in real data fusion, since the PSF and SRF may not be perfectly known in practice, and needs to be estimated in advance. Sim{\~{o}}es  \emph{et al.} \cite{hysure} introduce an approach  to estimate the PSF and SRF based on the following observation model
\begin{equation}\label{psfsrt}
\min_{{\bf P}_{3},{\bf G}}||{\bf P}_{3}{\bf Y}_{(3)}-{\bf Z}_{(3)}{\bf G}||_F^2+\lambda_1\Psi({\bf P}_{3})+\lambda_1\Phi({\bf G}),
\end{equation}
in which $\lambda_1\Psi({\bf P}_{3})$ and $\lambda_1\Phi({\bf G})$ denote the prior information on the ${\bf P}_{3}$ and ${\bf G}$, respectively.
They use the total variation to regularize the above two terms. However, they do not consider the non-negativity of the two terms and may obtain negative elements in ${\bf P}_{3}$ and ${\bf G}$. What is more, the PSF can often be modeled as a Gaussian blur kernel, and this strong prior information may further contribute to the estimation of PSF.\par

\subsection{Zero-short Learning for HSI-MSI Fusion}
The deep learning based fusion methods often take an end-to-end way to learn the mapping from low-resolution HSI and high-resolution MSI to fused HSI. The deep learning based methods can achieve promising performance in terms of quality of fused HSI and computation efficacy. However, methods in this category mainly have two disadvantages when applying them to real data fusion. Firstly, they suffer from insufficient training data. In practice, the training data for HSI-MSI fusion is often not available. Furthermore,  the well-trained deep CNN may have limited generalization ability, since the observational models, spectral range, and the number of spectral bands of the training data and testing data may be different.
The possible way to solve the problem is zero-short learning, which trains the deep CNN from the data to be fused. In the training procedure, the training data is generated by spatially downsampling the observed HSI and MSI, and the observed HSI is used as the output of deep CNN. In this way, the spatial downsampling procedure for generating training data makes a difference, and the way of spatially downsampling procedure should be learned according to the spatial degeneration between the observed HSI and MSI.

\subsection{Computation Efficiency}
Since the grown size of HSI and MSI data, the computation efficiency is a very important index for the HSI. In remote sensing HSI-MSI fusion, the spatial size is often very large, and the number of spectral bands is usually over one hundred. Hence, the fusion approaches with the low computational cost are highly favored.  The most of MF methods and TR methods suffer from high computational complexities since they need to iteratively solve the complex optimization problem. One available way to reduce the computational cost is low-dimensional subspace representation, which can significantly reduce the size of the spectral mode by exploiting the redundancies in the spectral mode. The other way is deep CNN based approaches, which take an end-to-end way to predict the fused HSI without iteration. What is more, they can be dramatically accelerated via the graphics processing unit (GPU).

\section{Conclusions}
This paper gives a comprehensive review for HSI-MSI fusion methods.
In specific, we classify the approaches into four main
families, i.e., pan-sharpening based  methods, MF based methods, TR based methods, and deep CNN based methods, and we give detailed introductions and discussions for approaches in each category. What is more, we also analyze the existing challenges for HSI-MSI fusion and present the new guidances and potential research directions for HSI-MSI fusion.
\section*{Acknowledgement}
This paper is supported by  the Major Program of the National Natural Science Foundation of China (No. 61890962), the National Natural Science Foundation of China (No. 61801178),  the Fund of Hunan Province for Science and Technology Plan Project under Grant (No. 2017RS3024),  the Fund of Key Laboratory of Visual Perception and Artificial Intelligence of Hunan Province (No. 2018TP1013), and the Natural Science Foundation of Hunan Province (No. 2019JJ50036).

\section*{References}
\bibliography{egbib}

\end{document}